\documentclass[11pt,a4paper]{article}
\usepackage[T1]{fontenc}
\usepackage{lmodern}
\usepackage[margin=1in]{geometry}
\usepackage{authblk}
\usepackage{graphicx}
\usepackage{amsmath,amssymb,amsfonts}
\usepackage{amsthm}
\usepackage[title]{appendix}
\usepackage{booktabs}
\usepackage{xcolor}
\usepackage{caption}
\usepackage{subcaption}
\usepackage{float}
\usepackage[authoryear,round]{natbib}
\usepackage{xurl}
\usepackage{listings}
\usepackage[hidelinks]{hyperref}
\usepackage{bookmark}
\graphicspath{{./}}
\setcitestyle{authoryear,round,aysep={},citesep={;}}
\hypersetup{
  pdftitle={Distributional Flexibility under Box-Cox Normality},
  pdfauthor={Kazushi Maruo, Yusuke Yamaguchi, Ryota Ishii, Toshio Shimokawwa, Masahiko Gosho}}
\newcommand{\orcid}[1]{\href{#1}{\textsuperscript{\normalfont\scriptsize iD}}}

\DeclareCaptionFont{supplementfont}{\fontsize{10}{12}\selectfont}
\newenvironment{tablenotes}[1][]{%
  \par\smallskip
  \begin{list}{}{%
    \setlength{\leftmargin}{0pt}%
    \setlength{\rightmargin}{0pt}%
    \setlength{\labelsep}{0pt}%
    \setlength{\labelwidth}{0pt}%
    \setlength{\topsep}{0pt}%
    \setlength{\itemsep}{0pt}%
    \setlength{\parsep}{0pt}%
  }%
}{\end{list}}

\newcommand{\continuedatlaspage}[2]{%
  \begin{figure}[p]
    \ContinuedFloat
    \centering
    \includegraphics[
      width=\textwidth,
      height=0.80\textheight,
      keepaspectratio
    ]{#1}
    \caption[]{#2}
  \end{figure}
  \clearpage
}

\title{Distributional Flexibility under Box--Cox Normality: A Theoretical and Empirical Evaluation of the Power-Normal Distribution}

\author[1]{Kazushi Maruo\orcid{https://orcid.org/0000-0002-3093-0386}%
\thanks{Correspondence: \href{mailto:kazushi.maruo@gmail.com}{kazushi.maruo@gmail.com}.}}
\author[2]{Yusuke Yamaguchi\orcid{https://orcid.org/0000-0002-8589-5303}}
\author[1]{Ryota Ishii\orcid{https://orcid.org/0000-0003-1725-4579}}
\author[3]{Toshio Shimokawa\orcid{https://orcid.org/0000-0002-2256-0282}}
\author[1]{Masahiko Gosho\orcid{https://orcid.org/0000-0002-5973-9163}}
\affil[1]{Department of Biostatistics, Institute of Medicine,
  University of Tsukuba, Tsukuba, Ibaraki, Japan}
\affil[2]{Biostatistics, Data Science,
  Astellas Pharma Global Development, Inc., Illinois, USA}
\affil[3]{Department of Biostatistics, Faculty of Medicine,
  Wakayama Medical University, Wakayama, Wakayama, Japan}
\date{}

\begin{document}

\maketitle

\begin{abstract}
Methods based on Box--Cox models have recently been developed for inference on clinically interpretable treatment effects in clinical trials.
These methods model marginal outcome distributions using the power-normal distribution (PND), but the range of real-world data distributions that it can adequately approximate remains unclear.
We evaluated PND approximations using two parameterizations of the four-parameter generalized $F$/generalized beta of the second kind family and 165 laboratory variables from the National Health and Nutrition Examination Survey.
Overall distributional agreement was assessed using the overlap coefficient and Jensen--Shannon divergence, and location and spread on the original scale were evaluated using relative biases in the median and interquartile range.
The PND closely approximated a broad range of smooth unimodal positive distributions in both the theoretical and empirical evaluations.
Reduced agreement was associated with pronounced lower-bound accumulation, heaping or discretization, and multicomponent or peak--tail features.
These features may be recognizable before fitting and provide practical guidance for assessing the suitability of the PND.
We also developed the R package \texttt{powerNormal} for distributional calculations, parameter estimation, reparameterization, and random-number generation for the PND.
\end{abstract}

\noindent\textbf{Keywords:} Model misspecification; Skewed data; Overlap coefficient; Clinical laboratory data

\clearpage

\section{Introduction}

Positively valued continuous outcomes with skewed distributions are
frequently encountered in biomedical and clinical research.
The Box--Cox transformation provides a classical and flexible framework for
analyzing such outcomes by transforming the data to a scale on which normality and homoscedasticity can be more reasonably assumed \citep{box_cox_64}.
This approach allows a broad range of distributional shapes to be accommodated
while retaining the use of well-established statistical methods based on
Gaussian models.
However, parameters and treatment effects defined on the transformed scale are
often difficult to interpret clinically.

To overcome this interpretability problem, methods have been developed that
combine the flexibility of Box--Cox transformation models with inference for
clinically interpretable treatment effects on the original measurement scale.
\citet{maruo_etal_15} proposed inference for differences in model medians between
treatment groups in randomized clinical trials, and
subsequently extended this framework to longitudinal data using mixed models
for repeated measures \citep{maruo_etal_17_bcmmrm}.
Both methods have been implemented in the R package \texttt{bcmixed}
\citep{maruo_etal_2021}.
More recently, this framework has been further extended to allow distributional
parameters to vary across treatment groups and measurement occasions, thereby
accommodating heterogeneous marginal distributional shapes
\citep{bcmvr}.
Under these Box--Cox-normal modeling approaches, the marginal distribution of a
positive outcome on the original scale is represented by the power-normal
distribution (PND), which is induced by assuming normality after a Box--Cox
transformation \citep{goto_etal_80,goto_etal_83}.
Thus, the ability of these methods to provide flexible yet interpretable
modeling on the original scale depends, at least in part, on how broadly the
PND can represent distributions encountered in practice.

The PND has long been regarded as a flexible family capable of representing a
wide variety of unimodal distributions, and its statistical properties,
parameterizations, and inferential procedures have been investigated in
previous studies.
Its empirical applicability has also been explored.
In particular, \citet{maruo_etal_17_cs} proposed a reparameterization of the PND using
interpretable quantities on the original scale and developed simulation
procedures for right-skewed data.
As part of that study, the PND was fitted to 113 laboratory measurements from
the 2009--2010 National Health and Nutrition Examination Survey (NHANES), and
goodness of fit in the central part of the distribution was assessed by
comparing PND-based and empirical medians.
Approximately 90\% of the ratios of the PND-based median to the sample median
were within 5\% of unity.
These findings provided broad empirical support for the ability of the PND to
reproduce the central location of right-skewed biomedical data.
However, the evaluation was primarily focused on the median and did not assess
agreement in distributional spread or in the overall density.

Although the flexibility of the PND is recognized, systematic evaluation
of its ability to reproduce distributional spread and the entire density
remains limited. Clarifying its approximation capability is important
both for understanding the PND itself and for assessing the applicability
of Box--Cox-normal models that perform inference on clinically interpretable treatment effects.

In this study, we evaluate the approximation capability of the PND from
theoretical and empirical perspectives. The theoretical evaluation uses
two complementary parameterizations of the four-parameter generalized
$F$/generalized beta distribution of the second kind family, which
represents a broad range of distributional shapes. The empirical
evaluation uses positive laboratory variables from the 2017--March 2020
pre-pandemic NHANES\@. Across both evaluations, we examine overall density
agreement and the reproduction of the median and interquartile range (IQR) to identify
distributional features associated with PND approximation performance.
We also provide the R package \texttt{powerNormal} \citep{pn}, which
implements distributional calculations and estimation for the PND.

\section{Power-normal distribution}

\subsection{Definition}
\label{subsec:pnd_definition}

Let $Y$ be a positive continuous random variable.
The Box--Cox transformation of $Y$ is defined as
\begin{equation*}
z(y;\lambda)
=
\begin{cases}
\dfrac{y^\lambda-1}{\lambda}, & \lambda \neq 0,\\[6pt]
\log y, & \lambda = 0,
\end{cases}
\end{equation*}
where $\lambda$ is the transformation parameter
\citep{box_cox_64}.
The power-normal distribution (PND) is defined as the distribution of
$Y$ induced by assuming normality for the Box--Cox transformed variable,
subject to the range of the transformation
\citep{goto_etal_80,goto_etal_83}.
Its probability density function is given by
\begin{equation}
f_{\mathrm{PN}}
(y;\lambda,\mu,\sigma)
=
\frac{
y^{\lambda-1}
}{
A\sqrt{2\pi}\sigma
}
\exp
\left[
-\frac{\{z(y;\lambda)-\mu\}^{2}}{2\sigma^{2}}
\right],
\qquad y>0,
\label{eq:pnd_pdf}
\end{equation}
where
\begin{equation*}
A =
\begin{cases}
\displaystyle
\Phi\!\left\{
\operatorname{sgn}(\lambda)
\frac{1+\lambda\mu}{\lambda\sigma}
\right\},
& \lambda \neq 0, \\[6pt]
1,
& \lambda = 0,
\end{cases}
\end{equation*}
and $\Phi(\cdot)$ denotes the cumulative distribution function of the
standard normal distribution.
The parameters $\lambda$, $\mu$, and $\sigma$ represent the shape,
location, and scale parameters, respectively. We denote the parameter vector by
$\boldsymbol{\theta}=(\lambda,\mu,\sigma)^\mathsf{T}$.

The range of the Box--Cox transformed variable depends on the sign of
$\lambda$.
For $\lambda>0$, $z(Y;\lambda)>-1/\lambda$, whereas for $\lambda<0$,
$z(Y;\lambda)<-1/\lambda$.
Thus, the normal distribution on the transformed scale is formally
truncated, and $A$ in Eq.~(\ref{eq:pnd_pdf}) is the corresponding
normalizing constant.
When $\lambda=1$, the PND corresponds to a normal distribution
truncated at zero, whereas $\lambda=0$ corresponds to the log-normal
distribution.
Negative values of $\lambda$ allow substantially heavier right tails
than the log-normal distribution.
Consequently, variation in $\lambda$ enables the PND to represent a
wide range of positively skewed unimodal distributional shapes.

Parameter estimation commonly assumes $A=1$, ignoring truncation on the
transformed scale. \citet{maruo_etal_11} investigated PND estimation with
the normalizing constant explicitly taken into account and showed that,
although this allows a broader range of distributions to be represented,
it can reduce estimation precision and cause convergence problems.
More complex Box--Cox models for original-scale inference therefore
assume $A=1$ to avoid these estimation difficulties and have nevertheless
shown good performance \citep{maruo_etal_15,maruo_etal_17_bcmmrm,bcmvr}.
Accordingly, we used the estimation criterion with $A=1$ in this study.

\subsection{Quantiles and original-scale parameterization}
\label{subsec:pnd_reparam}

Let $\xi_p$ denote the $100p$th percentile of the PND\@.
For the PND with the normalizing constant explicitly taken into
account, the quantile is given by
\begin{equation*}
\xi_p
=
\begin{cases}
\left[
1+\lambda\left\{\mu+\sigma z_{p^*}\right\}
\right]^{1/\lambda},
& \lambda\neq0,\\[6pt]
\exp(\mu+\sigma z_p),
& \lambda=0,
\end{cases}
\end{equation*}
where $z_p=\Phi^{-1}(p)$ and
\begin{equation*}
p^*
=
\begin{cases}
1-A(1-p), & \lambda>0,\\
Ap,       & \lambda<0.
\end{cases}
\end{equation*}
Under the approximation $A=1$, $p^*=p$, and the quantile function
reduces to the inverse Box--Cox transformation of the corresponding
normal quantile.

For applications in which interpretation on the original measurement
scale is important, the PND can be reparameterized as
$\boldsymbol{\omega}=\left(\lambda,\xi_{0.5},\tau\right)^\mathsf{T}$,
where $\xi_{0.5}$ is the median and serves as a location parameter
on the original measurement scale, and $\tau$ is a dimensionless
scale parameter \citep{maruo_goto_13,maruo_etal_17_cs}:
\begin{equation*}
\tau
=
\frac{\xi_{0.75}-\xi_{0.25}}
     {\xi_{0.5}}.
\end{equation*}
This parameterization provides directly interpretable location and
scale-related quantities on the original measurement scale.

The median and interquartile range used in the subsequent evaluations
were obtained from the corresponding quantile function.
All PND-related calculations were implemented in the R package
\texttt{powerNormal}, which provides distributional functions,
parameter estimation, reparameterization, and random-number
generation for the PND. A brief
description of its implementation and principal functions is provided
in Appendix~\ref{app:powernormal}.

\section{Theoretical evaluation of the approximation capability}

\subsection{Reference family and parameterizations}

To evaluate the approximation capability of the PND over a broad
range of positive continuous distributions, we used the generalized
$F$ (GF)/generalized beta distribution of the second kind (GB2) family as
the reference family. This four-parameter family accommodates wide
variation in skewness, peakedness, and tail behavior and contains or
approaches several familiar positive distributions as special or
limiting cases \citep{prentice_75,mcdonald_84,cox_08}.
Rather than treating the GF and GB2 distributions as
distinct reference families, we used them as complementary
parameterizations of the same underlying family.

First, we used the stable GF parameterization proposed by
\citet{prentice_75}. We denote its parameter vector by
$\boldsymbol{\eta}_1=(\gamma_1,\kappa_1,\tau_1,\nu_1)^\mathsf{T}$,
where $\gamma_1$ and $\kappa_1$ are shape parameters,
$\tau_1>0$ is a scale-related parameter, and $\nu_1$ is a
location parameter on the logarithmic scale. Define
\begin{equation*}
\delta
=
\left(\gamma_1^2+2\kappa_1\right)^{1/2},
\end{equation*}
\begin{equation*}
s_1
=
\frac{2}
{\gamma_1^2+2\kappa_1+\gamma_1\delta},
\qquad
s_2
=
\frac{2}
{\gamma_1^2+2\kappa_1-\gamma_1\delta},
\end{equation*}
and
\begin{equation*}
w
=
\frac{\delta(\log x-\nu_1)}{\tau_1}.
\end{equation*}
For $\kappa_1>0$, the density is
\begin{equation}
f_1(x)
=
\frac{
\delta
(s_1/s_2)^{s_1}
\exp(s_1 w)
}{
\tau_1 x B(s_1,s_2)
\left\{
1+(s_1/s_2)\exp(w)
\right\}^{s_1+s_2}
},
\qquad x>0,
\label{eq:gfd_pdf}
\end{equation}
where $B(\cdot,\cdot)$ denotes the beta function. At the boundary $\kappa_1=0$, Equation~\eqref{eq:gfd_pdf}
is interpreted by continuity; its limit yields the generalized
gamma family, including the log-normal distribution when
$\gamma_1=0$. Because the log-normal distribution is also
contained in the PND when $\lambda=0$, this setting is the only
correctly specified case among the reference distributions considered;
all other settings evaluate the approximation capability of the PND
under distributional misspecification.

We next used the GB2 parameterization of \citet{mcdonald_84}.
We denote its parameter vector by
$\boldsymbol{\eta}_2=(\gamma_2,\kappa_2,\tau_2,\nu_2)^\mathsf{T}$,
with all four parameters positive.
Its density is
\begin{equation*}
f_2(x)
=
\frac{
\tau_2
(x/\nu_2)^{\tau_2\gamma_2-1}
}{
\nu_2 B(\gamma_2,\kappa_2)
\left\{
1+(x/\nu_2)^{\tau_2}
\right\}^{\gamma_2+\kappa_2}
},
\qquad x>0.
\end{equation*}
Near the origin, the density is proportional to
$x^{\tau_2\gamma_2-1}$, whereas its right tail is proportional to
$x^{-(\tau_2\kappa_2+1)}$. Thus, $\gamma_2$ and $\kappa_2$ directly
govern behavior near the origin and in the right tail, respectively;
$\tau_2$ controls the power transformation and peakedness, and
$\nu_2$ is a scale parameter.

For $\kappa_1>0$, the GF and GB2 parameterizations are related by a
one-to-one transformation and represent the same four-parameter
family. The generalized gamma case at $\kappa_1=0$ forms a boundary
of this family. The GF parameterization facilitates interpretation
through familiar special and limiting distributions, whereas the
GB2 parameterization provides direct control of near-origin and
right-tail behavior. We therefore examined separate parameter grids
under the two parameterizations to obtain complementary coverage of
the same distributional family. Selected familiar special and
limiting cases are summarized in Table~\ref{tab:special_cases}.

The GF parameterization has been used mainly in parametric survival
analysis because it accommodates diverse distributional and hazard
shapes \citep{cox_08}. The GB2 parameterization has been used for income
and other positive size distributions with pronounced right-skewness
and heavy upper tails \citep{mcdonald_84}. Here, we used this family to
evaluate a broad range of positive distributional shapes rather than
to target a particular application area.

Numerical calculations under the GF and GB2 parameterizations were
performed using the R packages \texttt{flexsurv} and \texttt{GB2},
respectively \citep{flexsurv,gb2}.

\subsection{Fitting the PND under distributional misspecification}
Let $h(x;\boldsymbol{\eta})$ denote the density of a reference
distribution indexed by parameter vector $\boldsymbol{\eta}$. For each reference parameter setting, we generated a
Monte Carlo sample of size $n_{\mathrm{MC}}=10^6$ and fitted the PND
using the \texttt{pnd\_est} function in the R package \texttt{powerNormal}.
The resulting estimate,
\begin{equation*}
\widehat{\boldsymbol{\theta}}_{\boldsymbol{\eta}}
=
(\widehat{\lambda},
 \widehat{\mu},
 \widehat{\sigma})^\mathsf{T},
\end{equation*}
provides a numerical approximation to the population limiting
parameter
\begin{equation*}
\boldsymbol{\theta}_{\boldsymbol{\eta}}^{*}
=
\arg\max_{\boldsymbol{\theta}}
E_{\boldsymbol{\eta}}
\left[
\ell_{A=1}(X;\boldsymbol{\theta})
\right],
\end{equation*}
where $\ell_{A=1}(X;\boldsymbol{\theta})$ is the individual
Box--Cox-normal log-likelihood contribution used for estimation. Thus,
$\boldsymbol{\theta}_{\boldsymbol{\eta}}^{*}$ is the pseudo-true PND
parameter induced by the estimation criterion; it is not obtained by
directly optimizing any of the agreement measures defined below. The
large fitting sample approximates this population criterion while
avoiding potentially unstable direct integration for heavy-tailed
reference distributions.

\subsection{Measures of distributional agreement}
\label{subsec:agreement}

We used the overlap coefficient (OVL) as the primary measure of
overall distributional agreement \citep{inman_bradley_89}. For a
reference density $h(x;\boldsymbol{\eta})$ and the fitted PND density
$f_{\mathrm{PN}}(x;\boldsymbol{\theta}_{\boldsymbol{\eta}}^{*})$, the
OVL is defined as
\begin{equation}
\mathrm{OVL}
=
\int_0^\infty
\min
\left\{
h(x;\boldsymbol{\eta}),
f_{\mathrm{PN}}
(x;\boldsymbol{\theta}_{\boldsymbol{\eta}}^{*})
\right\}
dx.
\label{eq:ovl}
\end{equation}
The OVL ranges from 0 to 1, with larger values indicating greater
overlap and hence greater agreement between the two distributions.

As a complementary measure of overall distributional discrepancy, we
used the Jensen--Shannon divergence (JSD) \citep{lin_91}.
Let
\begin{equation*}
m(x)
=
\frac{
h(x;\boldsymbol{\eta})
+
f_{\mathrm{PN}}
(x;\boldsymbol{\theta}_{\boldsymbol{\eta}}^{*})
}{2}.
\end{equation*}
The JSD is defined as
\begin{equation}
\mathrm{JSD}
=
\frac{1}{2}
\mathrm{KL}\{h\,\|\,m\}
+
\frac{1}{2}
\mathrm{KL}\{f_{\mathrm{PN}}\,\|\,m\},
\label{eq:jsd}
\end{equation}
where $\mathrm{KL}\{f\,\|\,g\}$ denotes the Kullback--Leibler
divergence of $f$ relative to $g$, defined by
\begin{equation*}
\mathrm{KL}\{f\,\|\,g\}
=
\int_0^\infty
f(x)
\log_2
\left\{
\frac{f(x)}{g(x)}
\right\}
dx.
\end{equation*}
With logarithms to base 2, the JSD ranges from 0 to 1, with smaller
values indicating greater agreement.

The OVL and JSD were evaluated by Monte Carlo integration to assess
reference distributions with widely differing tail behavior using a
common procedure without relying on the choice of an integration grid.
In particular, the OVL can be written as
\begin{equation*}
\mathrm{OVL}
=
E_h
\left[
\min
\left\{
1,
\frac{
f_{\mathrm{PN}}
(X;\boldsymbol{\theta}_{\boldsymbol{\eta}}^{*})
}{
h(X;\boldsymbol{\eta})
}
\right\}
\right],
\end{equation*}
and was approximated using $n_{\mathrm{MC}}=10^6$ random draws from
the reference distribution. The JSD was evaluated from its two
Kullback--Leibler components using $10^6$ random draws from the
reference distribution and $10^6$ independent random draws from the
fitted PND\@.

In addition to overall distributional agreement, we evaluated
discrepancies in location and spread on the original measurement
scale.
Let $\xi_h(p)$ and $\xi_{\mathrm{PN}}(p)$ denote the $100p$th
percentiles of the reference distribution and fitted PND,
respectively.
The relative bias in the median was defined as
\begin{equation*}
B_{\mathrm{Med}}
=
100
\frac{
\xi_{\mathrm{PN}}(0.5)-\xi_h(0.5)
}{
\xi_h(0.5)
},
\end{equation*}
and the relative bias in the IQR was defined as
\begin{equation*}
B_{\mathrm{IQR}}
=
100
\frac{
\mathrm{IQR}_{\mathrm{PN}}-\mathrm{IQR}_{h}
}{
\mathrm{IQR}_{h}
},
\end{equation*}
where
\begin{equation*}
\mathrm{IQR}
=
\xi(0.75)-\xi(0.25).
\end{equation*}
Both biases are signed: positive values indicate that the fitted PND
overestimates the corresponding reference quantity, whereas negative
values indicate underestimation.

\subsection{Evaluation settings}

Under the GF parameterization, we considered
\begin{equation*}
\begin{aligned}
\tau_1 &\in \{0.25,0.5,1,2\},
&
\gamma_1 &\in \{-2,-1.9,\ldots,2\},
\\
\kappa_1 &\in \{0,0.1,\ldots,4\}.
\end{aligned}
\end{equation*}
Under the GB2 parameterization, we considered
\begin{equation*}
\begin{aligned}
\tau_2 &\in \{1,2,4\},
&
\gamma_2 &\in \{0.1,0.2,\ldots,5\},
\\
\kappa_2 &\in \{0.1,0.2,\ldots,5\}.
\end{aligned}
\end{equation*}

For each combination of the shape parameters, the location or scale
parameter $\nu_j$ was selected so that the median of the reference
distribution was 100.
Because all measures used for the present comparison are invariant to
a common rescaling of the distributions, fixing the median at 100
allowed distributional shapes to be compared on a common scale
without loss of generality.

For each setting, a PND was fitted as described above, and approximation
performance was summarized over the $(\gamma_j,\kappa_j)$ grid using
heat maps. For fixed values of the other shape parameters, changing
$\tau_j$ corresponds to a one-to-one power transformation of the
reference variable. The PND family is closed under power
transformations, and both the OVL and JSD are invariant when the same
one-to-one transformation is applied to two distributions. Therefore,
under the present fitting scheme, OVL and JSD do not depend on
$\tau_j$ and were evaluated only at the representative setting
$\tau_1=\tau_2=1$. In contrast, relative biases in the median and IQR
are not invariant on the original measurement scale and were evaluated
at all investigated values of $\tau_1$ and $\tau_2$.

Selected overlays of the reference and fitted PND densities were
constructed to examine distributional shapes in regions with reduced
agreement.

\begin{table}[t!]
\centering
\caption{Selected familiar distributions represented as special or
limiting cases of the GF/GB2 family under the two parameterizations}
\label{tab:special_cases}
\begin{tabular}{lll}
\toprule
Parameterization & Distribution & Parameter condition \\
\midrule
GF
  & Generalized gamma
  & $\kappa_1=0$ \\

  & Log-normal
  & $\kappa_1=0,\ \gamma_1=0$ \\

  & Weibull
  & $\kappa_1=0,\ \gamma_1=1$ \\

  & Inverse Weibull
  & $\kappa_1=0,\ \gamma_1=-1$ \\

  & Gamma
  & $\kappa_1=0,\ \gamma_1=\tau_1$ \\

  & Inverse gamma
  & $\kappa_1=0,\ \gamma_1=-\tau_1$ \\

  & Log-logistic
  & $\gamma_1=0,\ \kappa_1=1$ \\
\midrule
GB2
  & Beta prime
  & $\tau_2=1$ \\

  & Singh--Maddala (Burr XII)
  & $\gamma_2=1$ \\

  & Dagum (Burr III)
  & $\kappa_2=1$ \\

  & Log-logistic
  & $\gamma_2=1,\ \kappa_2=1$ \\

  & Lomax (Pareto II)
  & $\tau_2=1,\ \gamma_2=1$ \\
\bottomrule
\end{tabular}

\begin{tablenotes}[flushleft]
\footnotesize
\item[]
\textit{Note:}
Only selected cases useful for interpreting the parameter spaces
considered in this study are shown. The stated conditions concern the
shape parameters; the location/scale parameters $\nu_1$ and $\nu_2$
are omitted because they do not determine membership in the listed
subfamilies. Boundary or limiting interpretations apply where
appropriate.
\end{tablenotes}
\end{table}

\subsection{Results of the approximation performance evaluation}
\begin{figure}[b!]
\centering
\includegraphics[width=31pc]{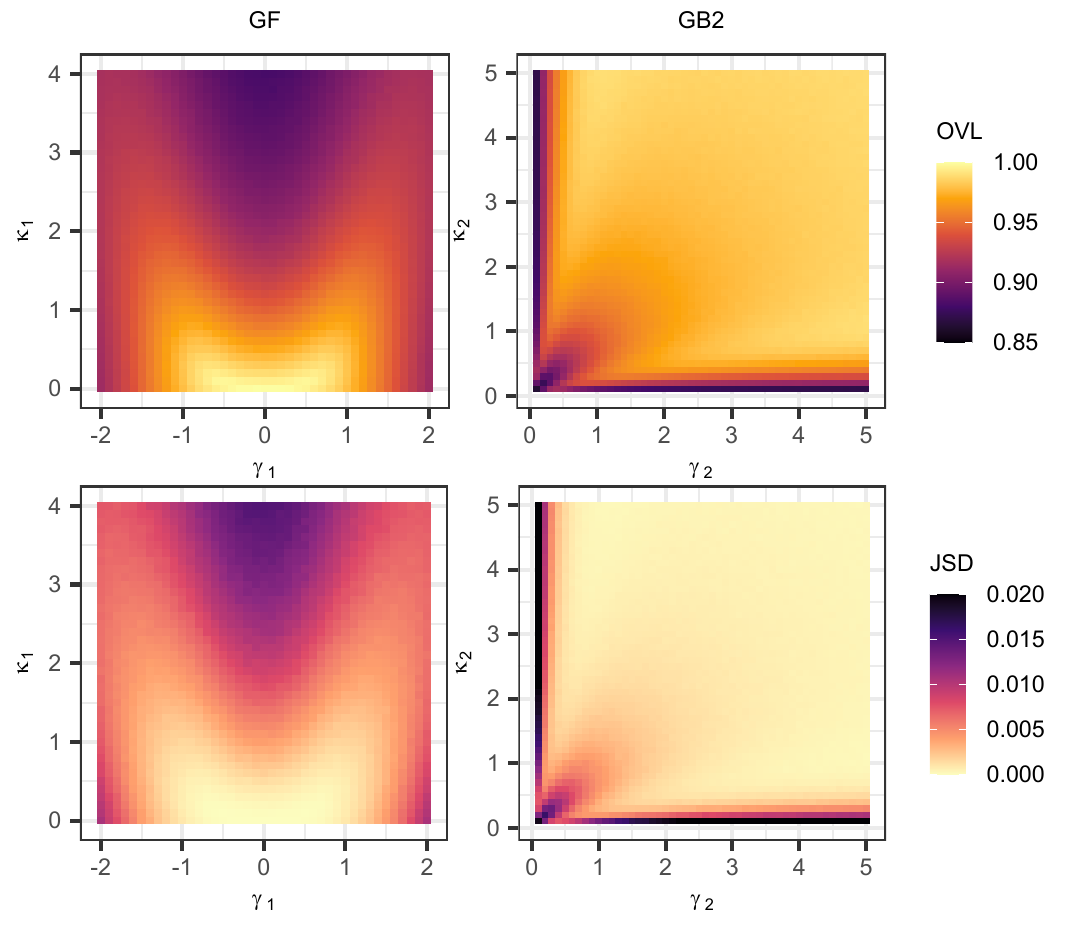}
\caption{Overall distributional agreement between each reference
distribution and its fitted power-normal distribution (PND). The top
and bottom rows show the overlap coefficient (OVL) and
Jensen--Shannon divergence (JSD), respectively; the left and right
columns correspond to the generalized $F$ (GF) and generalized beta
distribution of the second kind (GB2) parameterizations,
respectively. The horizontal and vertical axes are $\gamma_j$ and
$\kappa_j$, respectively. Results are shown at
$\tau_1=\tau_2=1$ because OVL and JSD are invariant to $\tau_j$ under
the present fitting scheme. JSD values above 0.02 are displayed using
the color corresponding to 0.02. Larger
OVL and smaller JSD indicate closer agreement}
\label{fig:overall_agreement}
\end{figure}

\begin{figure}[b!]
\centering
\includegraphics[width=31pc]{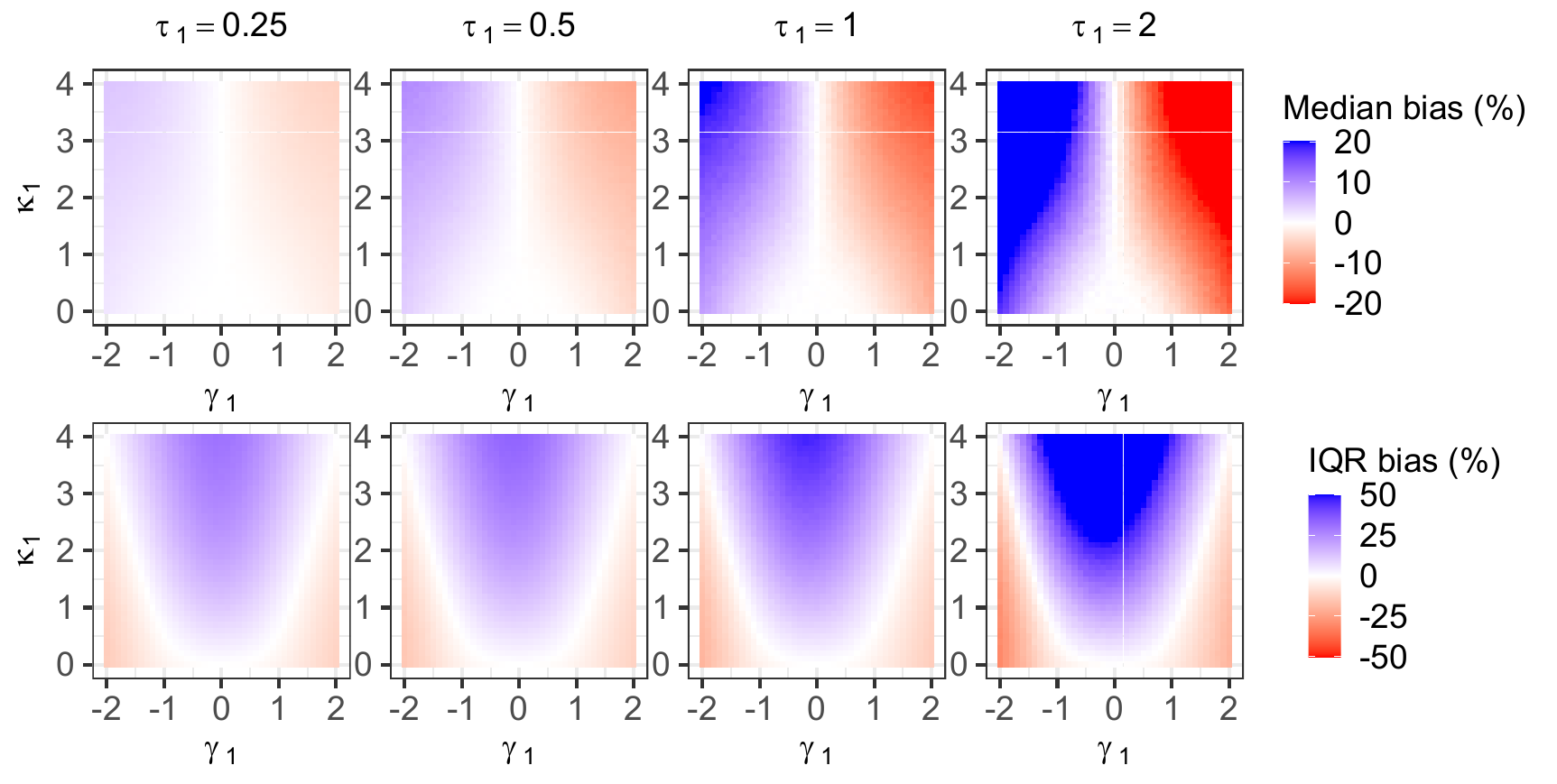}
\caption{Relative biases in the median and interquartile range (IQR)
of the fitted power-normal distribution (PND) under the generalized
$F$ (GF) parameterization. The top and bottom rows show median and IQR
relative biases (\%), respectively. From left to right, the columns
correspond to $\tau_1=0.25$, $0.5$, $1$, and $2$. The horizontal and
vertical axes are $\gamma_1$ and $\kappa_1$, respectively. Positive
values indicate overestimation by the PND and negative values indicate
underestimation. Median biases outside $[-20,20]\%$ and IQR biases
outside $[-50,50]\%$ are displayed at the corresponding color-scale
limits}
\label{fig:genf_bias}
\end{figure}

\begin{figure}[b!]
\centering
\includegraphics[width=31pc]{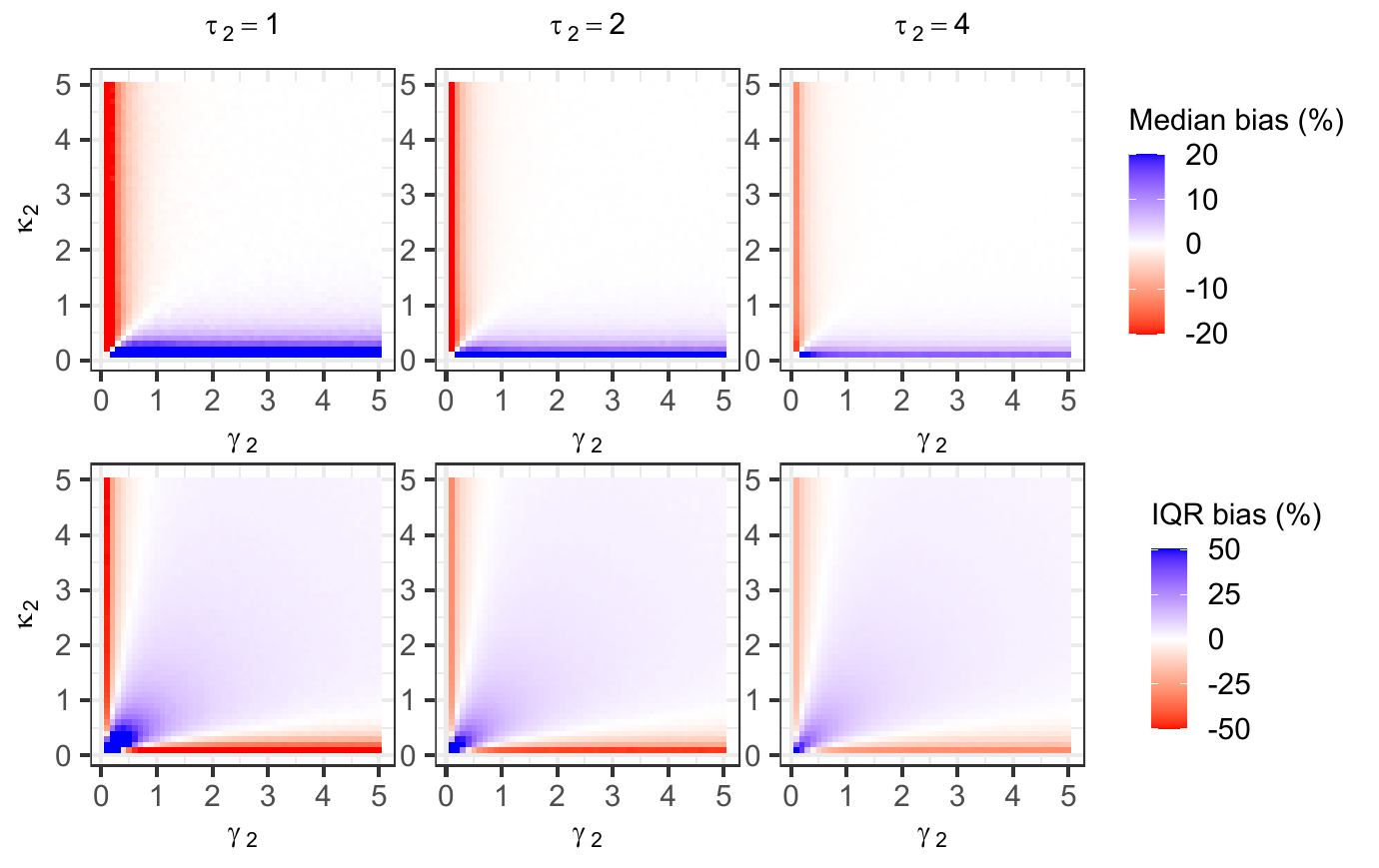}
\caption{Relative biases in the median and interquartile range (IQR)
of the fitted power-normal distribution (PND) under the generalized
beta distribution of the second kind (GB2) parameterization. The top
and bottom rows show median and IQR relative biases (\%), respectively.
From left to right, the columns correspond to $\tau_2=1$, $2$, and
$4$. The horizontal and vertical axes are $\gamma_2$ and $\kappa_2$,
respectively. Positive values indicate overestimation by the PND and
negative values indicate underestimation. Median biases outside
$[-20,20]\%$ and IQR biases outside $[-50,50]\%$ are displayed at the
corresponding color-scale limits}
\label{fig:gb2_bias}
\end{figure}
Figure~\ref{fig:overall_agreement} shows the overall distributional
agreement between the reference distributions and their fitted PNDs.
Across the 4,181 settings evaluated at $\tau_1=\tau_2=1$, the OVL
ranged from 0.864 to nearly 1 and the JSD from 0 to 0.0239. Regions
with lower OVL closely coincided with those with higher JSD\@.
Agreement was relatively lower when $\kappa_1$ was large and
$\gamma_1$ was close to zero under the GF parameterization, and when
either $\gamma_2$ or $\kappa_2$ was small under the GB2
parameterization, particularly when both were small.

The density overlays in Supplementary Figures~S1 and~S2 clarify the
shapes associated with reduced agreement. Small $\gamma_2$,
corresponding primarily to the $\gamma_1>0$ side of the GF
parameterization, produced pronounced concentration near zero. Small
$\kappa_2$, corresponding primarily to the $\gamma_1<0$ side,
produced a relatively narrow main probability mass accompanied by a
very long right tail. When both $\gamma_2$ and $\kappa_2$ were small, a sharp main peak was combined with substantial boundary or
tail features.

Figures~\ref{fig:genf_bias} and~\ref{fig:gb2_bias} summarize the
relative biases in the median and IQR\@. Under the GF parameterization,
the median bias was approximately zero near $\gamma_1=0$, positive
when $\gamma_1<0$, and negative when $\gamma_1>0$, with its absolute
magnitude generally increasing with $|\gamma_1|$ and $\tau_1$. Under
the GB2 parameterization, the median bias was negative when
$\gamma_2$ was small and positive when $\kappa_2$ was small. When both
features were present, these opposing effects could partially cancel,
so that the median bias remained small even when the OVL and JSD
indicated reduced overall agreement.

The IQR bias more closely reflected overall distributional agreement.
Under the GF parameterization, it tended to be positive when
$\kappa_1$ was large and $\gamma_1$ was close to zero, and negative
when $|\gamma_1|$ was large. Under the GB2 parameterization, large
absolute IQR biases occurred mainly when $\gamma_2$ or $\kappa_2$ was
small, particularly when both were small, with the direction of bias
varying where the two features competed. Absolute quantile biases
generally increased with larger $\tau_1$ and smaller $\tau_2$.
Because the OVL and JSD are invariant to $\tau_j$, these changes
reflect amplification of location and spread discrepancies on the
original measurement scale rather than changes in overall
distributional agreement.

Discrepancies in the IQR or overall density were observed even when
median bias was small. Approximation performance should therefore be
assessed in terms of spread and overall distributional agreement as
well as the median.

\section{Empirical evaluation using NHANES laboratory data}
\label{sec:nhanes}

\subsection{Data source and selection of laboratory variables}
\label{subsec:nhanes_data}

We used data from the Laboratory component of the 2017--March 2020
pre-pandemic National Health and Nutrition Examination Survey
(NHANES). The data were obtained using version 1.4.1 of the R package
\texttt{nhanesA} \citep{nhanesA}. All 51 Laboratory component
tables listed for this survey period were successfully retrieved.

Because our objective was to evaluate the observed sample distributions
rather than to estimate distributions representative of the US population,
NHANES survey sampling weights were not used.

After participant identifiers and survey-weight variables were
excluded, all non-empty numeric variables in the Laboratory component
tables were treated as candidates. Variables were excluded if they
had fewer than 100 non-missing observations, fewer than 20 distinct
observed values, any non-positive observations, or an empirical IQR of
zero. When the same measurement was available in multiple units, one
variable was retained according to a prespecified unit priority.
The final analysis included 165 variables, with variable-specific
sample sizes ranging from 1,835 to 12,510. These included variables
related to laboratory measurements, such as urine volume and
collection intervals. Details of the variable-screening flow and unit
harmonization are provided in the analysis code.

\subsection{Statistical analysis}
\label{subsec:nhanes_analysis}

A PND was fitted to each variable. Relative biases in the median and
IQR, denoted by $B_{\mathrm{Med}}^{\mathrm{emp}}$ and
$B_{\mathrm{IQR}}^{\mathrm{emp}}$, respectively, were calculated as in
Section~\ref{subsec:agreement}, using empirical quantiles as the
reference.

The empirical density was obtained by applying Gaussian kernel density
estimation to the log-transformed observations. The bandwidth was
selected separately for each variable using Silverman's rule of thumb
\citep{silverman_86}.

Because the empirical and fitted PND densities can be evaluated on a
common grid on the logarithmic scale, the OVL and JSD in
Eqs.~(\ref{eq:ovl}) and~(\ref{eq:jsd}) were calculated by numerical
integration without additional Monte Carlo error. These measures are
invariant under a common one-to-one transformation, so the resulting
values are the same as those obtained by comparing the corresponding
densities on the original scale. Implementation details and
computational settings are provided in the analysis code.

We conducted an exploratory descriptive examination of variables with
relatively low agreement to identify common features of their
empirical distributions. Empirical histograms and kernel density
estimates were compared with the fitted PND densities on both the
original and logarithmic scales.

\subsection{Empirical approximation performance}
\label{subsec:nhanes_results}

PND fitting and distributional evaluation were completed successfully
for all 165 variables. Variable-specific parameter estimates and
performance measures are provided in Supplementary Table~S1.

\begin{figure}[b!]
  \centering
  \includegraphics[width=31pc]
  {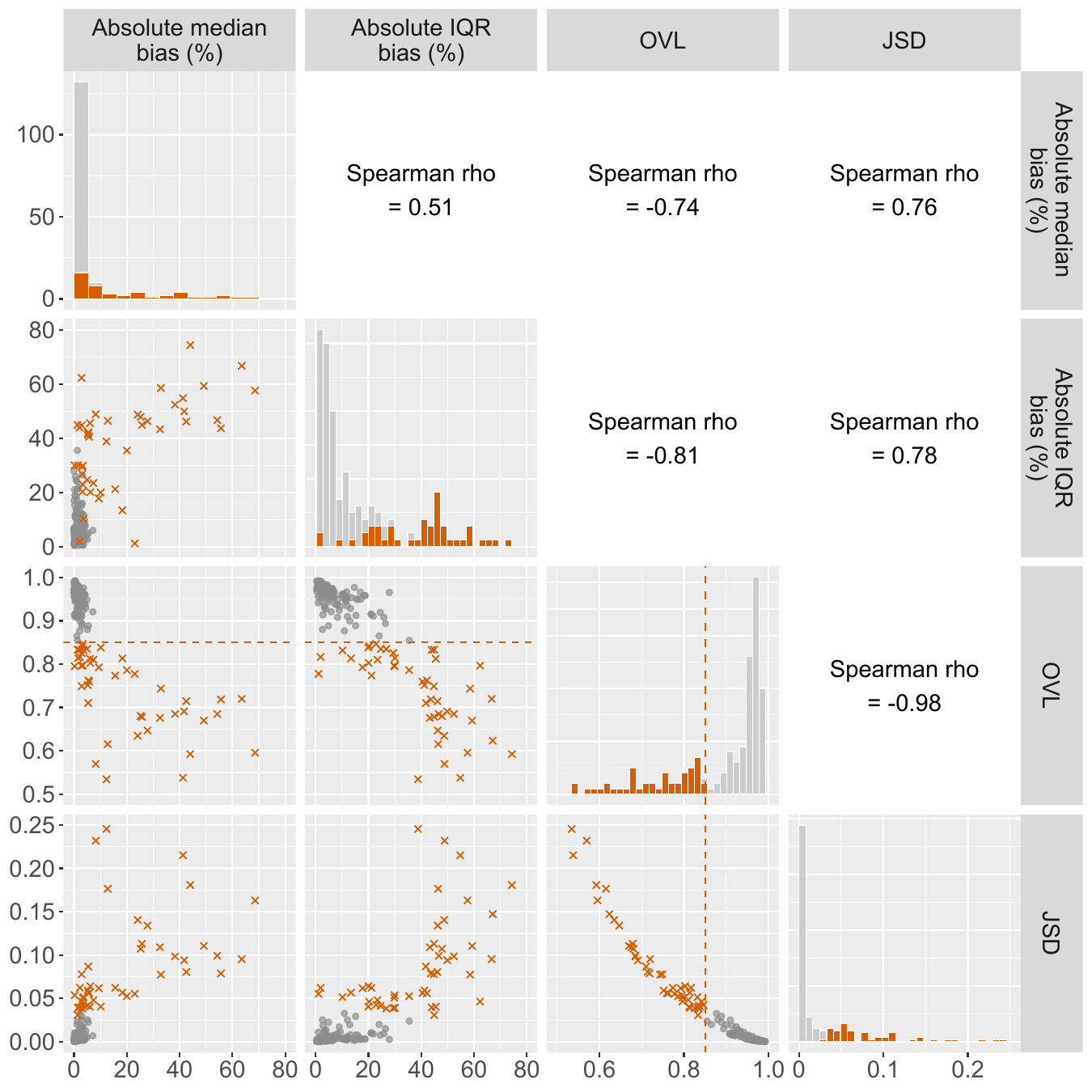}
  \caption{Empirical approximation performance of the fitted PND across 165 NHANES laboratory variables. Diagonal panels show the distributions of the absolute relative median bias, absolute relative IQR bias, OVL, and JSD; vermilion bars identify variables with OVL$<$0.85. Lower-triangle panels show pairwise relationships, with these variables indicated by vermilion crosses. Vermilion dashed lines indicate an OVL of 0.85. Upper-triangle panels report Spearman rank correlation coefficients. For visualization, hydroxycotinine (absolute relative median bias, 161.6\%) is omitted from the relevant histogram and scatterplots; summaries and correlations include all 165 variables.}
  \label{fig:nhanes_performance}
\end{figure}

Across the 165 variables, the median OVL was 0.949 (25th--75th
percentiles, 0.833--0.968), and the median JSD was 0.0044
(0.0020--0.0404). The median absolute relative biases were 1.8\% for
the median and 8.1\% for the IQR\@. The distributions and pairwise
relationships of the four performance measures are shown in
Figure~\ref{fig:nhanes_performance}, and numerical summaries are
presented in Table~\ref{tab:nhanes_performance}.

\begin{table}[t!]
\centering
\caption{Empirical approximation performance overall, by OVL-defined
agreement group, and by distributional pattern in the lower-agreement group}
\label{tab:nhanes_performance}
\fontsize{8}{10}\selectfont
\setlength{\tabcolsep}{2pt}
\renewcommand{\arraystretch}{1.15}

\begin{tabular}{lccccc}
\toprule
\multicolumn{6}{l}{
\textit{Panel A. Overall and by OVL-defined agreement group}
} \\
\addlinespace[2pt]

Group
& \shortstack{$n$\\(\%)}
& OVL
& JSD
& \shortstack{$\lvert B_{\mathrm{Med}}^{\mathrm{emp}}\rvert$\\(\%)}
& \shortstack{$\lvert B_{\mathrm{IQR}}^{\mathrm{emp}}\rvert$\\(\%)} \\
\midrule

All variables
& \shortstack{165\\(100.0)}
& \shortstack{0.949\\(0.833, 0.968)}
& \shortstack{0.0044\\(0.0020, 0.0404)}
& \shortstack{1.8\\(0.5, 4.2)}
& \shortstack{8.1\\(3.7, 24.1)} \\

$\mathrm{OVL}<0.85$
& \shortstack{47\\(28.5)}
& \shortstack{0.760\\(0.679, 0.813)}
& \shortstack{0.062\\(0.049, 0.108)}
& \shortstack{10.2\\(3.5, 32.8)}
& \shortstack{43.3\\(25.7, 48.3)} \\

$\mathrm{OVL}\geq0.85$
& \shortstack{118\\(71.5)}
& \shortstack{0.964\\(0.945, 0.974)}
& \shortstack{0.0028\\(0.0016, 0.0061)}
& \shortstack{1.0\\(0.4, 2.5)}
& \shortstack{5.0\\(2.8, 10.8)} \\

\addlinespace[6pt]
\multicolumn{6}{l}{
\shortstack[l]{\textit{Panel B. Distributional patterns identified among the}\\\textit{47 variables with $\mathrm{OVL}<0.85$}}
} \\
\addlinespace[2pt]

Pattern
& \shortstack{$n$\\(\%)}
& OVL
& JSD
& \shortstack{$B_{\mathrm{Med}}^{\mathrm{emp}}$\\(\%)}
& \shortstack{$B_{\mathrm{IQR}}^{\mathrm{emp}}$\\(\%)} \\
\midrule

\shortstack[l]{Lower-bound\\accumulation}
& \shortstack{32\\(68.1)}
& \shortstack{0.701\\(0.644, 0.770)}
& \shortstack{0.095\\(0.055, 0.136)}
& \shortstack{18.4\\($-0.5$, 41.9)}
& \shortstack{$-45.5$\\($-50.5$, $-36.6$)} \\

\shortstack[l]{Heaping/\\discretization}
& \shortstack{6\\(12.8)}
& \shortstack{0.808\\(0.795, 0.816)}
& \shortstack{0.062\\(0.058, 0.064)}
& \shortstack{$-2.7$\\($-5.5$, 1.6)}
& \shortstack{$-18.9$\\($-20.2$, $-14.5$)} \\

\shortstack[l]{Multicomponent/\\peak--tail pattern}
& \shortstack{9\\(19.1)}
& \shortstack{0.810\\(0.786, 0.832)}
& \shortstack{0.048\\(0.040, 0.053)}
& \shortstack{1.4\\($-2.9$, 7.4)}
& \shortstack{35.4\\(21.2, 44.9)} \\
\bottomrule
\end{tabular}

\begin{tablenotes}[flushleft]
\footnotesize
\item[]
\textit{Note:}
Values are medians (25th, 75th percentiles) unless otherwise indicated.
Percentages in Panel A use all 165 variables as the denominator; percentages
in Panel B use the 47 variables with $\mathrm{OVL}<0.85$. Panel A summarizes
absolute relative biases, whereas Panel B summarizes signed relative biases
to show the direction of error. The distributional patterns in Panel B were
identified through the exploratory visual examination described in the text.
\end{tablenotes}
\end{table}

The OVL and JSD produced nearly identical rankings of approximation quality
(Spearman $\rho=-0.981$). Lower OVL values were also associated with larger
absolute median and IQR biases ($\rho=-0.738$ and $-0.806$, respectively).
In contrast, the correlation between the absolute median and IQR biases was
moderate ($\rho=0.510$).

In the theoretical evaluation, the OVL was at least 0.864 in every
setting examined. We therefore focused on empirical distributional
shapes with lower agreement than those observed in that evaluation.
For this purpose, we selected the 47 variables (28.5\%) with OVL
below 0.85, an exploratory cutoff within the observed gap between
empirical OVL values of 0.846 and 0.855. Performance summaries for
both groups are shown in Table~\ref{tab:nhanes_performance}.

The distributions of these 47 variables were visually examined on the
original and logarithmic scales and classified into the following
three patterns according to their dominant features (Supplementary
Figures~S3 and~S4).

The most frequent pattern was lower-bound accumulation, characterized by a
pronounced concentration at the smallest observed values and a distinct
remainder of the distribution. This pattern was observed in 32 variables
(68.1\%). All 18 variables for which the empirical first quartile equaled the
empirical median belonged to this group. The median signed IQR bias was
$-45.5$\%, and 31 of the 32 variables had negative IQR biases, indicating a
strong tendency toward underestimation of spread. This group also had the
largest median absolute median bias among the three patterns, at 24.7\%.

Heaping or discretization, characterized by concentration at repeated values
and a comb-like empirical density, was the dominant pattern in 6 variables
(12.8\%). This group also included accumulation at finite upper reporting
limits in the poliovirus variables. All 6 variables had negative IQR biases,
with a median signed bias of $-18.9$\%. In contrast, the median absolute
median bias was 4.7\%, indicating that the median was comparatively well
reproduced despite the lack of agreement in the density shape.

The remaining 9 variables (19.1\%) showed a multicomponent or peak--tail
pattern, including multiple modes or shoulders, or a sharp central peak
accompanied by a broad tail or a remote component. All 9 variables had
positive IQR biases, with a median signed bias of 35.4\%. The median absolute
median bias was 6.0\%, and the median signed median bias was 1.4\%,
indicating that substantial overestimation of spread could occur even when
the median was reproduced relatively well.

\section{Discussion}
\label{sec:discussion}

This study systematically evaluated the approximation capability of the PND using the flexible four-parameter GF/GB2 family and 165 NHANES variables. The OVL was at least 0.864 in every theoretical setting examined, demonstrating that the three-parameter PND approximates a broad range of distributional shapes with high agreement. Good approximations were also obtained for many empirical variables, and the evaluation identified characteristic distributional shapes associated with reduced agreement.

This study extends the previous evaluation, which focused mainly on agreement between empirical and PND-based medians \citep{maruo_etal_17_cs}, to the IQR and the entire density. Accurate reproduction of the median did not necessarily imply close agreement in spread or distributional shape. The OVL and JSD provided nearly identical assessments and were more strongly associated with absolute IQR bias than with absolute median bias. These findings demonstrate the value of assessing approximation performance through spread and the entire density as well as central location.

In the theoretical evaluation, agreement was reduced for distributions with pronounced concentration near the lower boundary or a sharp main peak combined with extreme tails. The fitted PND had difficulty reproducing central concentration and tail behavior simultaneously in these settings. In addition, relative biases in the median and IQR varied under power transformations, whereas the OVL and JSD were invariant. This finding highlights the value of assessing the reproduction of quantities of interest on the original scale in addition to overall density agreement.

In NHANES, lower-bound accumulation, heaping/discretization, and multicomponent or peak--tail patterns were associated with reduced agreement. The direction of IQR discrepancy also differed across patterns: spread was generally underestimated in the first two and overestimated in the third. These features can be recognized in distribution plots before fitting and can help assess the suitability of the PND.

Box--Cox-normal methods for cross-sectional, longitudinal, and distributional regression analyses \citep{maruo_etal_15,maruo_etal_17_bcmmrm,bcmvr} use the PND to define clinically interpretable parameters on the original scale. This study characterizes the range of PND approximations underlying these methods and identifies distributional shapes for which fit warrants closer examination. Distribution plots on both the original and logarithmic scales are useful for this assessment.

The present evaluation covered the investigated GF/GB2 parameter ranges and the positive NHANES variables satisfying the selection criteria. Future work should examine a broader range of distributional shapes.

The PND closely approximated a broad range of smooth unimodal positive distributions using few parameters. By assessing the median, spread, and entire density together, this study clarifies the approximation capability of the PND and identifies distributional shapes for which fit warrants closer examination.

\subsection*{Supplementary information}
Supplementary Figures S1--S4, Supplementary Table S1, and the analysis
scripts accompany this article.

\section*{Statements and Declarations}

\subsection*{Funding}
This work was supported by the Japan Society for the Promotion of Science
(JSPS) KAKENHI under Grant Numbers 23K11003 and 26K02873.

\subsection*{Competing interests}
The authors declare that they have no competing interests relevant to this work.

\subsection*{Ethics approval and consent to participate}
This study used publicly available NHANES data and did not involve recruitment
of participants or collection of new participant data.

\subsection*{Consent for publication}
Not applicable.

\subsection*{Data availability}
The NHANES data analyzed in this study are publicly available from the US
National Center for Health Statistics through the
\href{https://wwwn.cdc.gov/nchs/nhanes/continuousnhanes/default.aspx?Cycle=2017-2020}{2017--March 2020 pre-pandemic data portal}.
Variable-specific parameter estimates and performance measures for the
165 variables analyzed in this study are provided in Supplementary Table S1.

\subsection*{Code availability}
The R scripts used for the theoretical and NHANES analyses are provided
as supplementary material.
The \texttt{powerNormal} R package and its documentation are publicly
available at \mbox{\url{https://github.com/kzkzmr/powerNormal}}.
The package version and commit used in this study are specified in the
reference list.


\begin{appendices}

\section{Implementation in the \textnormal{\texttt{powerNormal}} R package}
\label{app:powernormal}

The open-source R package \texttt{powerNormal} implements distributional
and estimation functions for the PND\@. The functions \texttt{dpnd},
\texttt{ppnd}, \texttt{qpnd}, and \texttt{rpnd} provide the density,
distribution, quantile, and random-generation functions, respectively,
under the conventional parameterization in terms of the shape, location,
and scale parameters $(\lambda,\mu,\sigma)$.

As described in Section~\ref{subsec:pnd_reparam}, functions whose names
end in \texttt{\_rp}---\texttt{dpnd\_rp}, \texttt{ppnd\_rp},
\texttt{qpnd\_rp}, and \texttt{rpnd\_rp}---use the original-scale
parameterization $(\lambda,\xi_{0.5},\tau)$ in place of
$(\lambda,\mu,\sigma)$. The function \texttt{pnd\_reparm} converts the
former parameterization to the latter before the distributional functions
are evaluated.

The function \texttt{pnd\_est} estimates $(\lambda,\mu,\sigma)$ from a
positive-valued sample. We used its default option, \texttt{tn = FALSE},
in this study (Section~\ref{subsec:pnd_definition}).

{\raggedright
The package and its documentation are publicly available at
\mbox{\url{https://github.com/kzkzmr/powerNormal}}. A minimal example is:\par}

\begin{lstlisting}
devtools::install_github("kzkzmr/powerNormal")
library(powerNormal)

set.seed(12345)

# Generate a PND with median 100 and IQR/median equal to 1

x <- rpnd_rp(
   n = 200,
   lambda = 0.5,
   xi = 100,
   tau = 1
)

# Estimate lambda, mu, and sigma

fit <- pnd_est(x, tn = FALSE)

# Quartiles of the fitted PND

qhat <- qpnd(
   c(0.25, 0.50, 0.75),
   fit$lambda,
   fit$mu,
   fit$sigma
)
\end{lstlisting}

\end{appendices}

\bibliographystyle{plainnat}
\bibliography{bibfile}

@article{box_cox_64,
  author  = {G. E. P. Box and D. R. Cox},
  title   = {An Analysis of Transformations},
  journal = {Journal of the Royal Statistical Society, Series B},
  year    = {1964},
  volume  = {26},
  number  = {2},
  pages   = {211--252},
  doi     = {10.1111/j.2517-6161.1964.tb00553.x}
}

@article{maruo_etal_17_bcmmrm,
  author  = {K. Maruo and Y. Yamaguchi and H. Noma and M. Gosho},
  title   = {Interpretable inference on the mixed effect model with the {Box}--{Cox} transformation},
  journal = {Statistics in Medicine},
  year    = {2017},
  volume  = {36},
  number  = {15},
  pages   = {2420--2434},
  doi     = {10.1002/sim.7279}
}

@article{maruo_etal_15,
  author  = {K. Maruo and N. Isogawa and M. Gosho},
  title   = {Inference of median difference based on the {Box}--{Cox} model in randomized clinical trials},
  journal = {Statistics in Medicine},
  year    = {2015},
  volume  = {34},
  number  = {10},
  pages   = {1634--1644},
  doi     = {10.1002/sim.6408}
}

@article{maruo_etal_17_cs,
  author  = {K. Maruo and T. Yamabe and Y. Yamaguchi},
  title   = {Statistical simulation based on right skewed distributions},
  journal = {Computational Statistics},
  year    = {2017},
  volume  = {32},
  pages   = {889--907},
  doi     = {10.1007/s00180-016-0664-4}
}

@article{goto_etal_80,
  author  = {M. Goto and T. Inoue},
  title   = {Some properties of the power normal distribution},
  journal = {Japanese Journal of Biometrics},
  year    = {1980},
  volume  = {1},
  pages   = {28--54},
  doi     = {10.5691/jjb.1.28}
}

@article{goto_etal_83,
  author  = {M. Goto and M. Matsubara and Y. Tsuchiya},
  title   = {Power-normal distribution and its applications},
  journal = {Reports of Statistical Application Research, JUSE},
  year    = {1983},
  volume  = {30},
  pages   = {8--28}
}

@article{maruo_etal_2021,
  author  = {K. Maruo and R. Ishii and Y. Yamaguchi and M. Gosho},
  title   = {bcmixed: {A} Package for Median Inference on Longitudinal Data with the {Box}--{Cox} Transformation},
  journal = {The R Journal},
  year    = {2021},
  volume  = {13},
  number  = {2},
  pages   = {253--265},
  doi     = {10.32614/RJ-2021-083},
  issn    = {2073-4859}
}

@article{maruo_etal_11,
  author  = {Maruo, Kazushi and Shirahata, Shingo and Goto, Masashi},
  title   = {Underlying Assumptions of the Power-Normal Distribution},
  journal = {Behaviormetrika},
  year    = {2011},
  volume  = {38},
  pages   = {85--95},
  doi     = {10.2333/bhmk.38.85}
}

@article{maruo_goto_13,
  author  = {Maruo, Kazushi and Goto, Masashi},
  title   = {Percentile Estimation Based on the Power-Normal Distribution},
  journal = {Computational Statistics},
  year    = {2013},
  volume  = {28},
  pages   = {341--356},
  doi     = {10.1007/s00180-012-0303-7}
}

@article{prentice_75,
  author  = {Prentice, R. L.},
  title   = {Discrimination among some parametric models},
  journal = {Biometrika},
  year    = {1975},
  volume  = {62},
  number  = {3},
  pages   = {607--614},
  doi     = {10.1093/biomet/62.3.607}
}

@article{mcdonald_84,
  author  = {McDonald, James B.},
  title   = {Some Generalized Functions for the Size Distribution of Income},
  journal = {Econometrica},
  year    = {1984},
  volume  = {52},
  number  = {3},
  pages   = {647--663}
}

@article{cox_08,
  author  = {Cox, Christopher},
  title   = {The Generalized {F} Distribution: An Umbrella for Parametric Survival Analysis},
  journal = {Statistics in Medicine},
  year    = {2008},
  volume  = {27},
  number  = {21},
  pages   = {4301--4312},
  doi     = {10.1002/sim.3292}
}

@article{inman_bradley_89,
  author  = {Inman, Henry F. and Bradley, {~Jr.}, Edwin L.},
  title   = {The Overlapping Coefficient as a Measure of Agreement between Probability Distributions and Point Estimation of the Overlap of Two Normal Densities},
  journal = {Communications in Statistics - Theory and Methods},
  year    = {1989},
  volume  = {18},
  number  = {10},
  pages   = {3851--3874},
  doi     = {10.1080/03610928908830127}
}

@article{lin_91,
  author  = {Lin, Jianhua},
  title   = {Divergence Measures Based on the {Shannon} Entropy},
  journal = {IEEE Transactions on Information Theory},
  year    = {1991},
  volume  = {37},
  number  = {1},
  pages   = {145--151},
  doi     = {10.1109/18.61115}
}

@Article{flexsurv,
    title = {{flexsurv}: A Platform for Parametric Survival Modeling in {R}},
    author = {Christopher Jackson},
    journal = {Journal of Statistical Software},
    year = {2016},
    volume = {70},
    number = {8},
    pages = {1--33},
    doi = {10.18637/jss.v070.i08}
  }

@Manual{gb2,
    title = {{GB2}: Generalized Beta Distribution of the Second Kind: Properties,
Likelihood, Estimation},
    author = {Monique Graf and Desislava Nedyalkova},
    year = {2025},
    note = {{R} package version 2.1.2},
    url = {https://CRAN.R-project.org/package=GB2},
    doi = {10.32614/CRAN.package.GB2}
  }

@article{nhanesA,
  title   = {{nhanesA}: achieving transparency and reproducibility
             in {NHANES} research},
  author  = {Ale, Laha and
             Gentleman, Robert and
             Filshtein Sonmez, Teresa and
             Sarkar, Deepayan and
             Endres, Christopher},
  journal = {Database},
  year    = {2024},
  volume  = {2024},
  pages   = {baae028},
  doi     = {10.1093/database/baae028}
}

@Manual{pn,
    title = {{powerNormal}: The power normal distribution with or without reparametrization},
    author = {Kazushi Maruo},
    year = {2026},
    note = {{R} package version 0.1.1, commit 173a14dd217c8ac293d8e03957989b385307e81a},
    url = {https://github.com/kzkzmr/powerNormal},
  }

@book{silverman_86,
  author    = {Silverman, Bernard W.},
  title     = {Density Estimation for Statistics and Data Analysis},
  year      = {1986},
  publisher = {Chapman and Hall},
  address   = {London}
}

@misc{bcmvr,
  author = {K. Maruo and R. Ishii and Y. Yamaguchi and
            T. Shimokawa and T. Sugimoto and M. Gosho},
  title = {A flexible framework for treatment effect inference in
           longitudinal clinical studies with skewed outcomes},
  howpublished = {arXiv preprint arXiv:2609.16670},
  year = {2026},
  doi = {10.48550/arXiv.2609.16670}
}

\clearpage
\newgeometry{left=1in,right=1in,top=1in,bottom=1in,bindingoffset=0pt}
\pagestyle{plain}
\fontsize{11}{13.6}\selectfont
\setlength{\parindent}{0pt}
\setlength{\parskip}{1\baselineskip}
\setcounter{figure}{0}
\setcounter{table}{0}
\renewcommand{\thefigure}{S\arabic{figure}}
\renewcommand{\thetable}{S\arabic{table}}
\renewcommand{\theHfigure}{supp.\arabic{figure}}
\renewcommand{\theHtable}{supp.\arabic{table}}
\renewcommand{\thesubfigure}{\alph{subfigure}}
\captionsetup{
  font=supplementfont,
  labelfont=bf,
  labelsep=colon,
  justification=justified,
  singlelinecheck=false
}
\captionsetup[figure]{name=Figure}
\captionsetup[subfigure]{
  font=supplementfont,
  labelfont=normal,
  justification=centering,
  singlelinecheck=true
}
\section*{Supplementary material}


\begin{figure}[H]
  \centering
  \setcounter{subfigure}{0}
  \begin{subfigure}{0.78\textwidth}
    \centering
    \includegraphics[width=\linewidth]{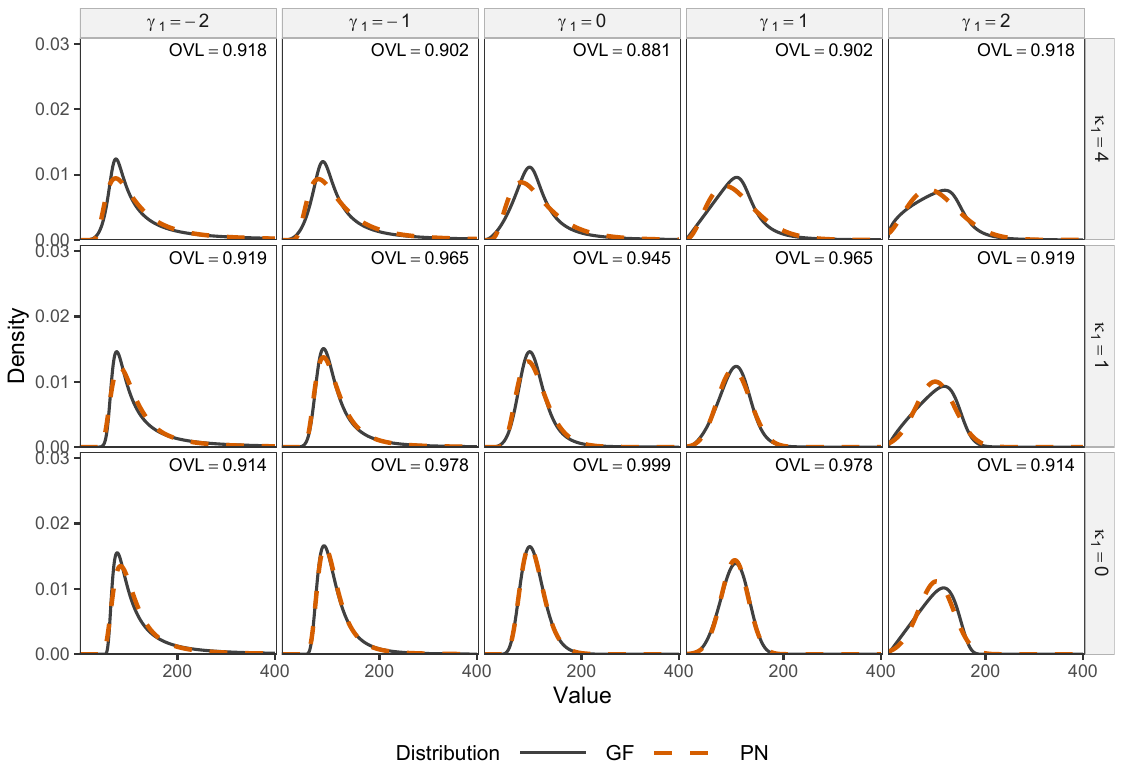}
    \caption{$\tau_1=0.25$}
    \label{fig:s1a}
  \end{subfigure}

  \vspace{0.15\baselineskip}

  \begin{subfigure}{0.78\textwidth}
    \centering
    \includegraphics[width=\linewidth]{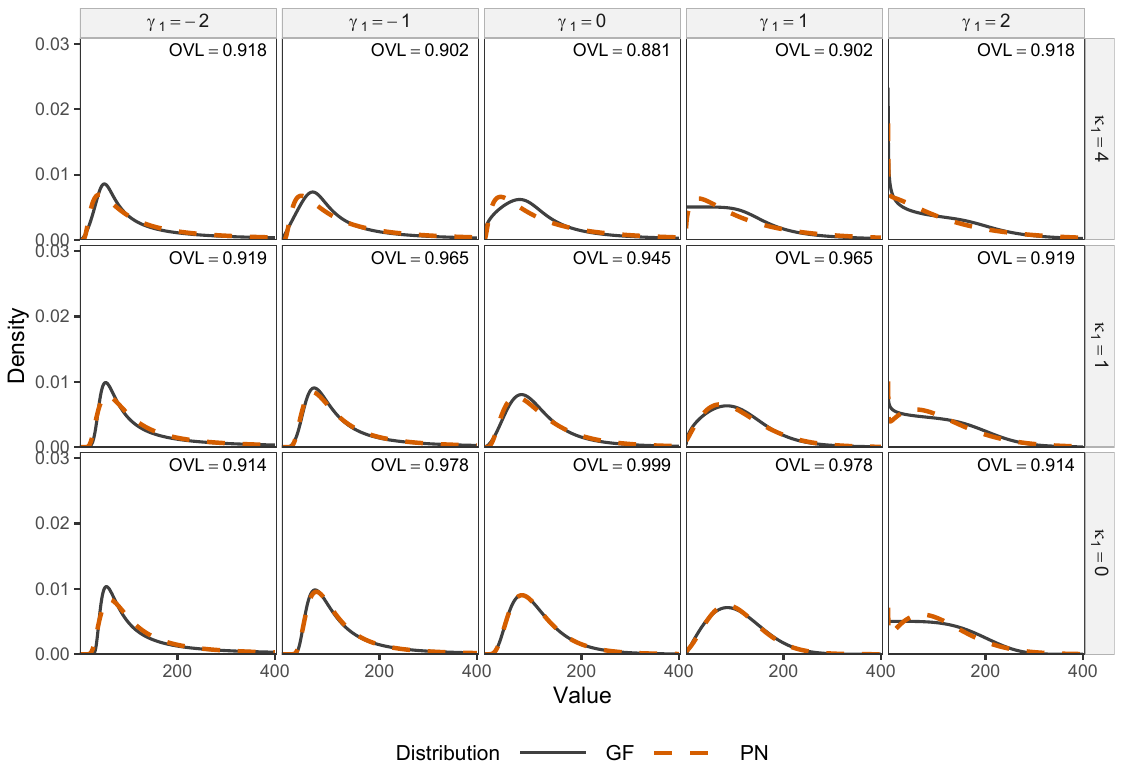}
    \caption{$\tau_1=0.5$}
    \label{fig:s1b}
  \end{subfigure}
  \caption{Comparison of reference densities under the GF parameterization
  with fitted PND densities. Panels (a)--(d) correspond to $\tau_1=0.25$,
  $0.5$, $1$, and $2$, respectively. Rows and columns correspond to the displayed
  values of $\kappa_1$ and $\gamma_1$, respectively. Dark-gray solid and
  vermilion dashed curves show the reference and fitted PND densities,
  respectively; all reference distributions were rescaled to a median of 100.}
  \label{fig:s1}
\end{figure}

\clearpage

\begin{figure}[p]
  \ContinuedFloat
  \centering
  \setcounter{subfigure}{2}
  \begin{subfigure}{0.84\textwidth}
    \centering
    \includegraphics[width=\linewidth]{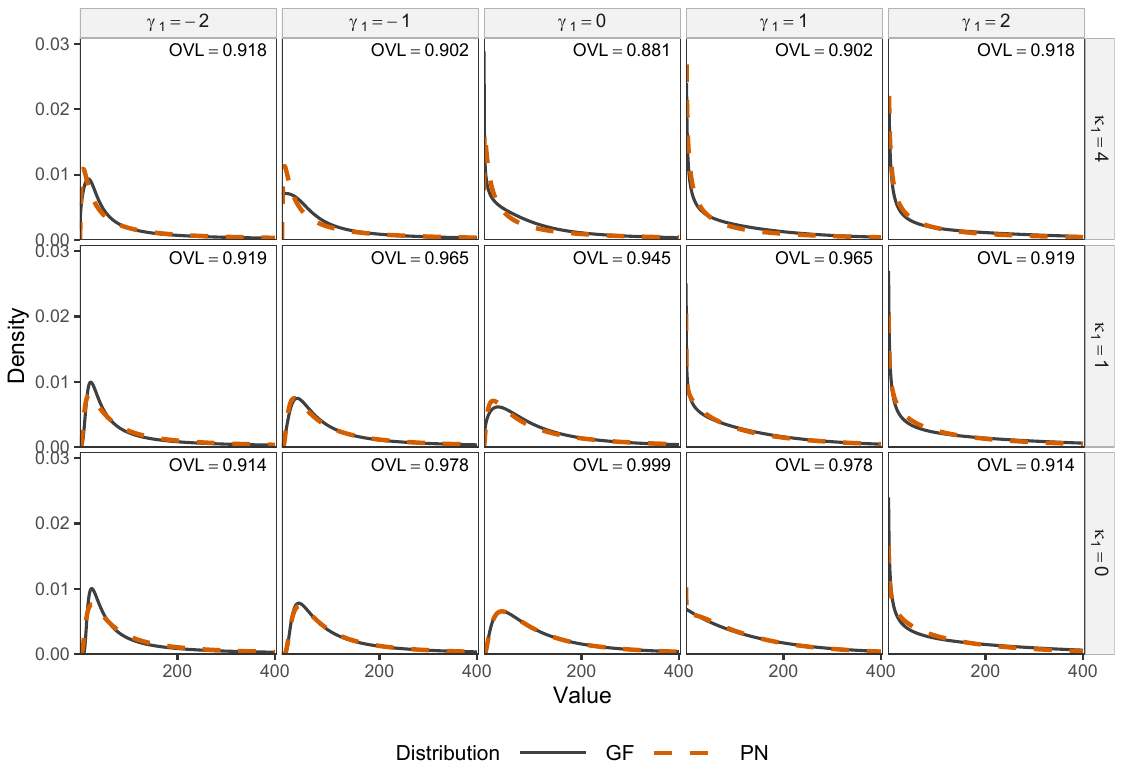}
    \caption{$\tau_1=1$}
    \label{fig:s1c}
  \end{subfigure}

  \vspace{0.15\baselineskip}

  \begin{subfigure}{0.84\textwidth}
    \centering
    \includegraphics[width=\linewidth]{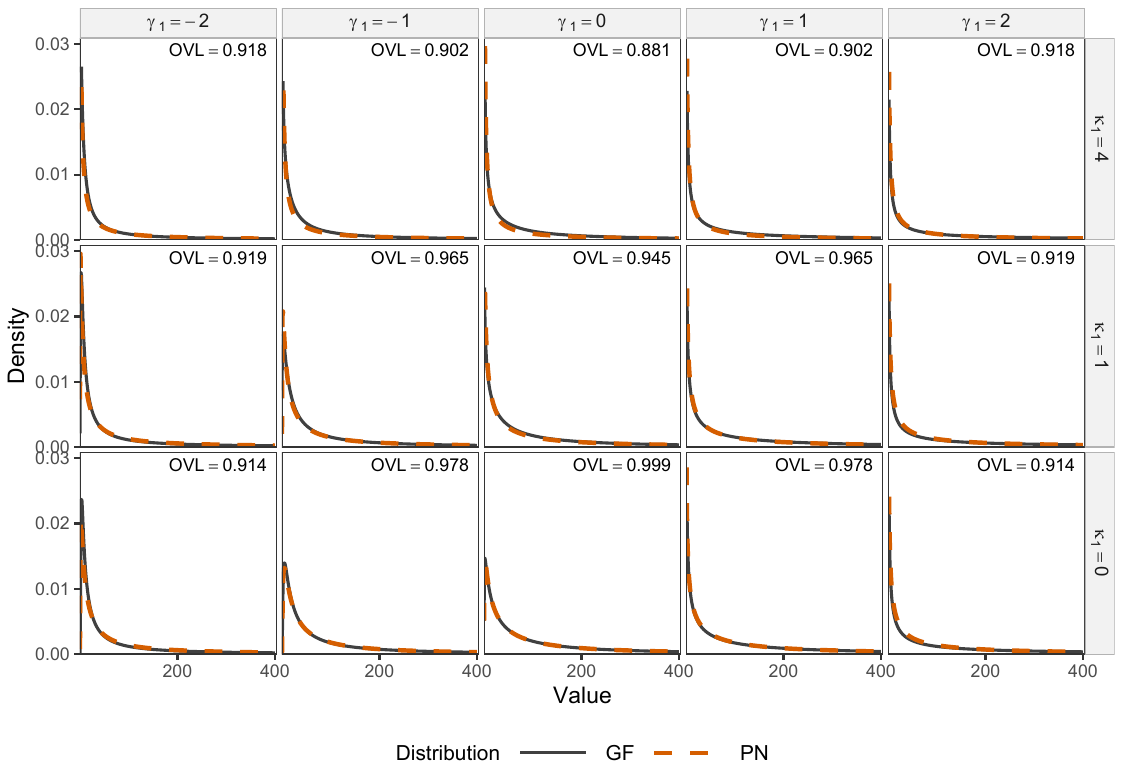}
    \caption{$\tau_1=2$}
    \label{fig:s1d}
  \end{subfigure}
  \caption[]{Comparison of reference GF and fitted PND densities
  (continued).}
\end{figure}

\clearpage


\begin{figure}[p]
  \centering
  \setcounter{subfigure}{0}
  \begin{subfigure}{0.84\textwidth}
    \centering
    \includegraphics[width=\linewidth]{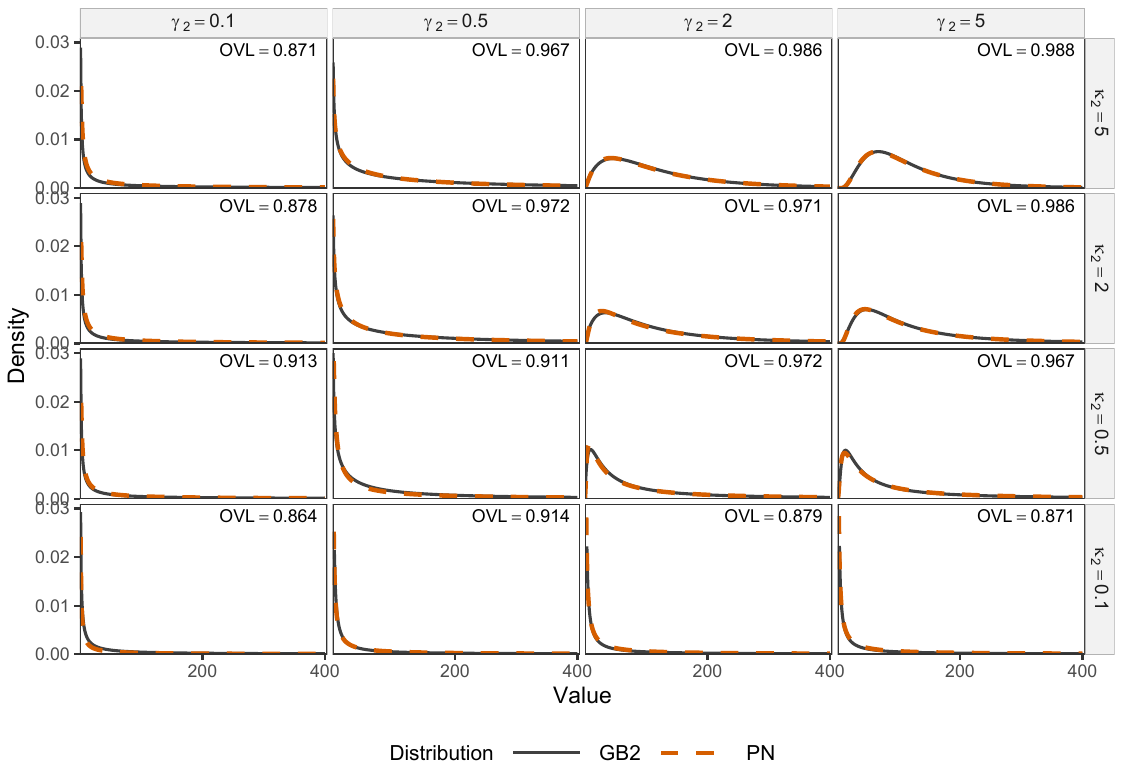}
    \caption{$\tau_2=1$}
    \label{fig:s2a}
  \end{subfigure}

  \vspace{0.15\baselineskip}

  \begin{subfigure}{0.84\textwidth}
    \centering
    \includegraphics[width=\linewidth]{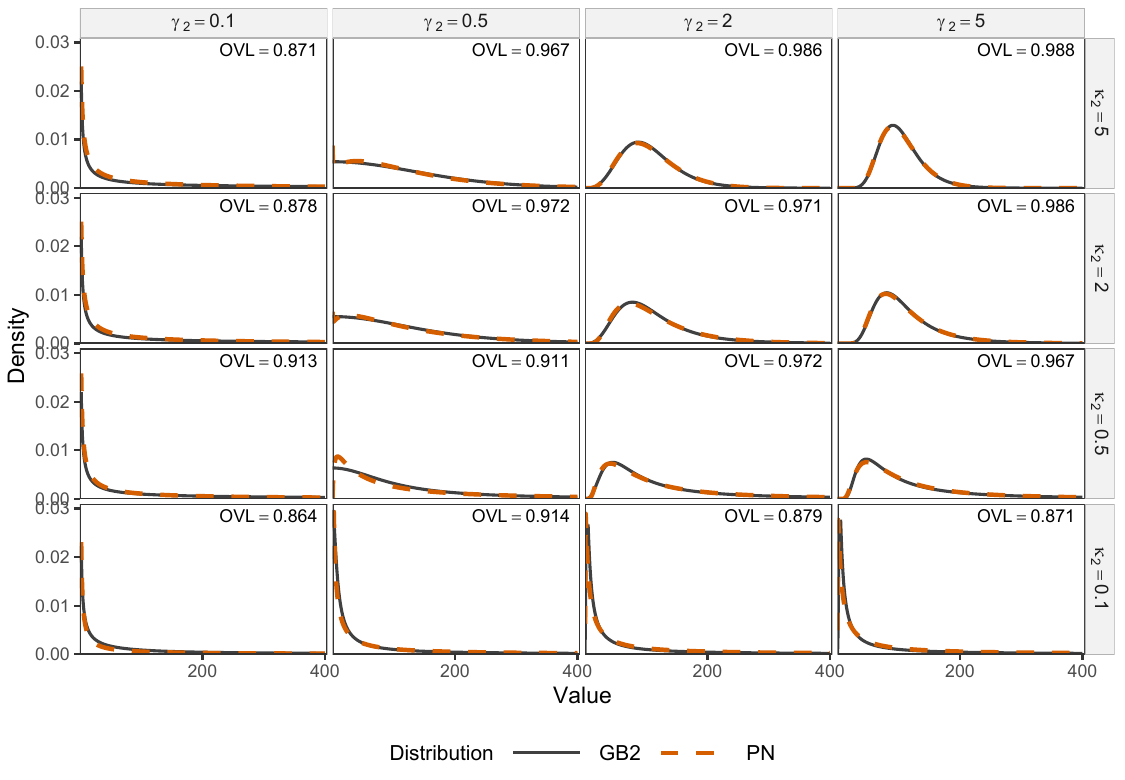}
    \caption{$\tau_2=2$}
    \label{fig:s2b}
  \end{subfigure}
  \caption{Comparison of reference densities under the GB2 parameterization
  with fitted PND densities. Panels (a)--(c) correspond to $\tau_2=1$, $2$, and
  $4$, respectively. Rows and columns correspond to the displayed values of
  $\kappa_2$ and $\gamma_2$, respectively. Line styles and rescaling are as in
  Figure~\ref{fig:s1}.}
  \label{fig:s2}
\end{figure}

\clearpage

\begin{figure}[p]
  \ContinuedFloat
  \centering
  \setcounter{subfigure}{2}
  \begin{subfigure}{0.84\textwidth}
    \centering
    \includegraphics[width=\linewidth]{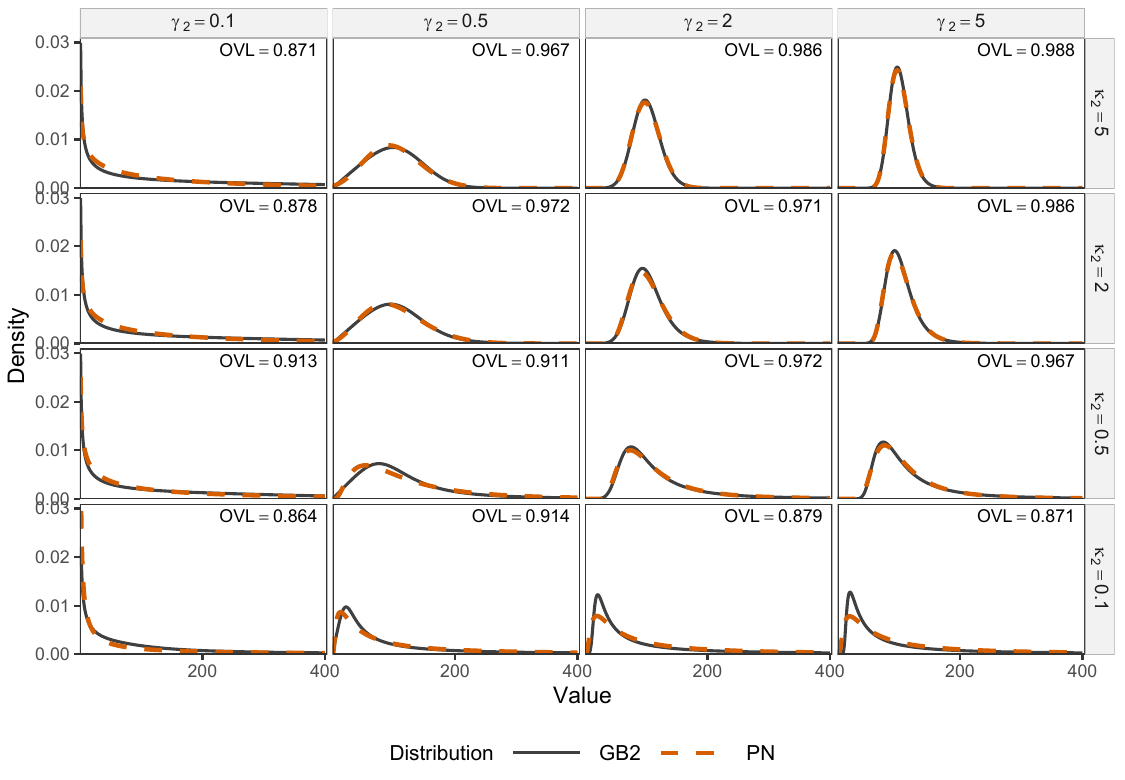}
    \caption{$\tau_2=4$}
    \label{fig:s2c}
  \end{subfigure}
  \caption[]{Comparison of reference GB2 and fitted PND densities
  (continued).}
\end{figure}

\clearpage


\begin{figure}[p]
  \centering
  \includegraphics[
    width=\textwidth,
    height=0.80\textheight,
    keepaspectratio
  ]{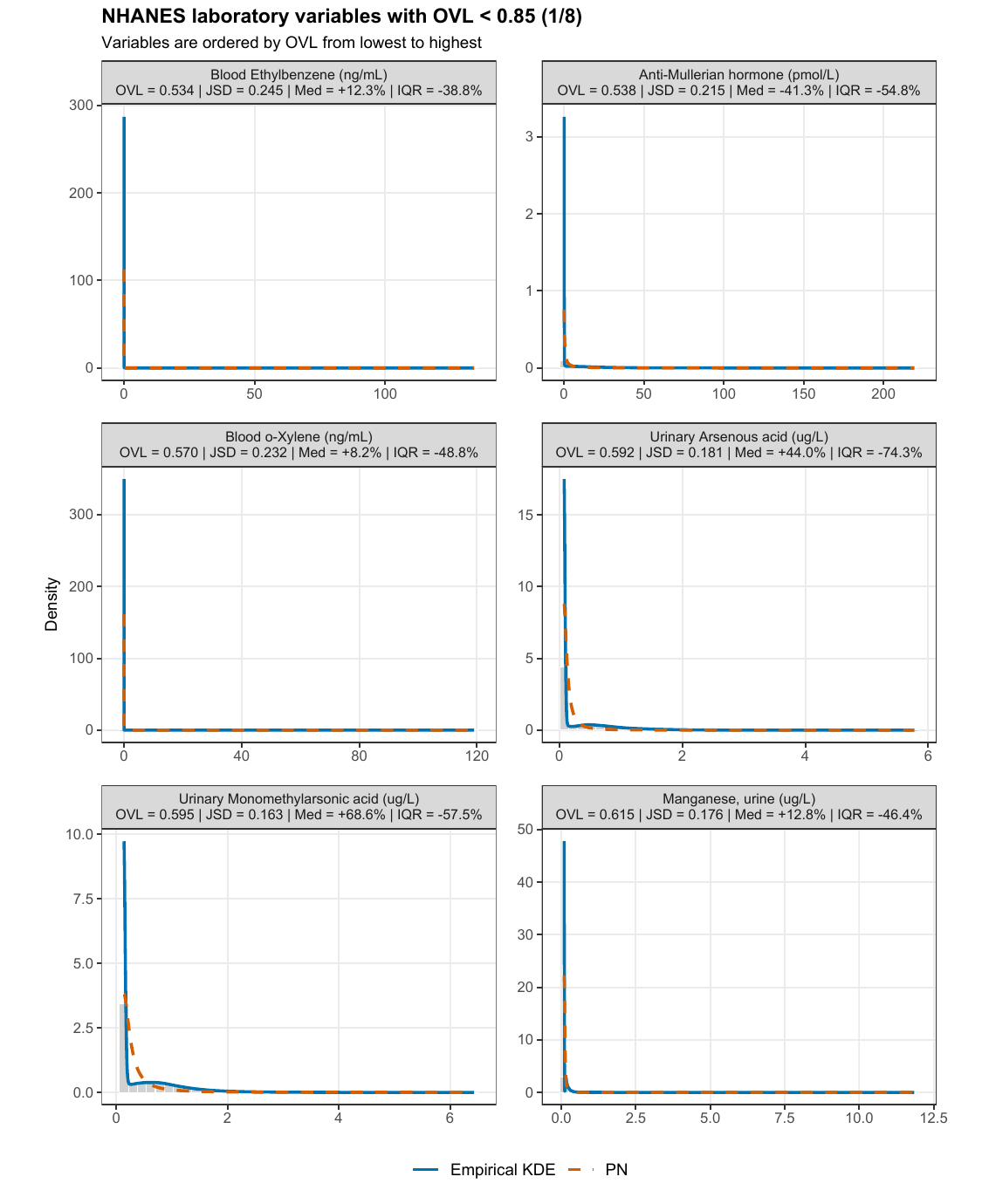}
  \caption{Empirical distributions and fitted PND densities on the original
  measurement scale for the 47 NHANES laboratory variables with
  $\mathrm{OVL}<0.85$. Variables are ordered from the lowest to the highest
  OVL. Gray bars show the empirical histogram, the blue solid curve shows the
  Gaussian kernel density estimate, and the vermilion dashed curve shows the
  fitted PND density. All observations were retained.}
  \label{fig:s3}
\end{figure}
\clearpage

\continuedatlaspage{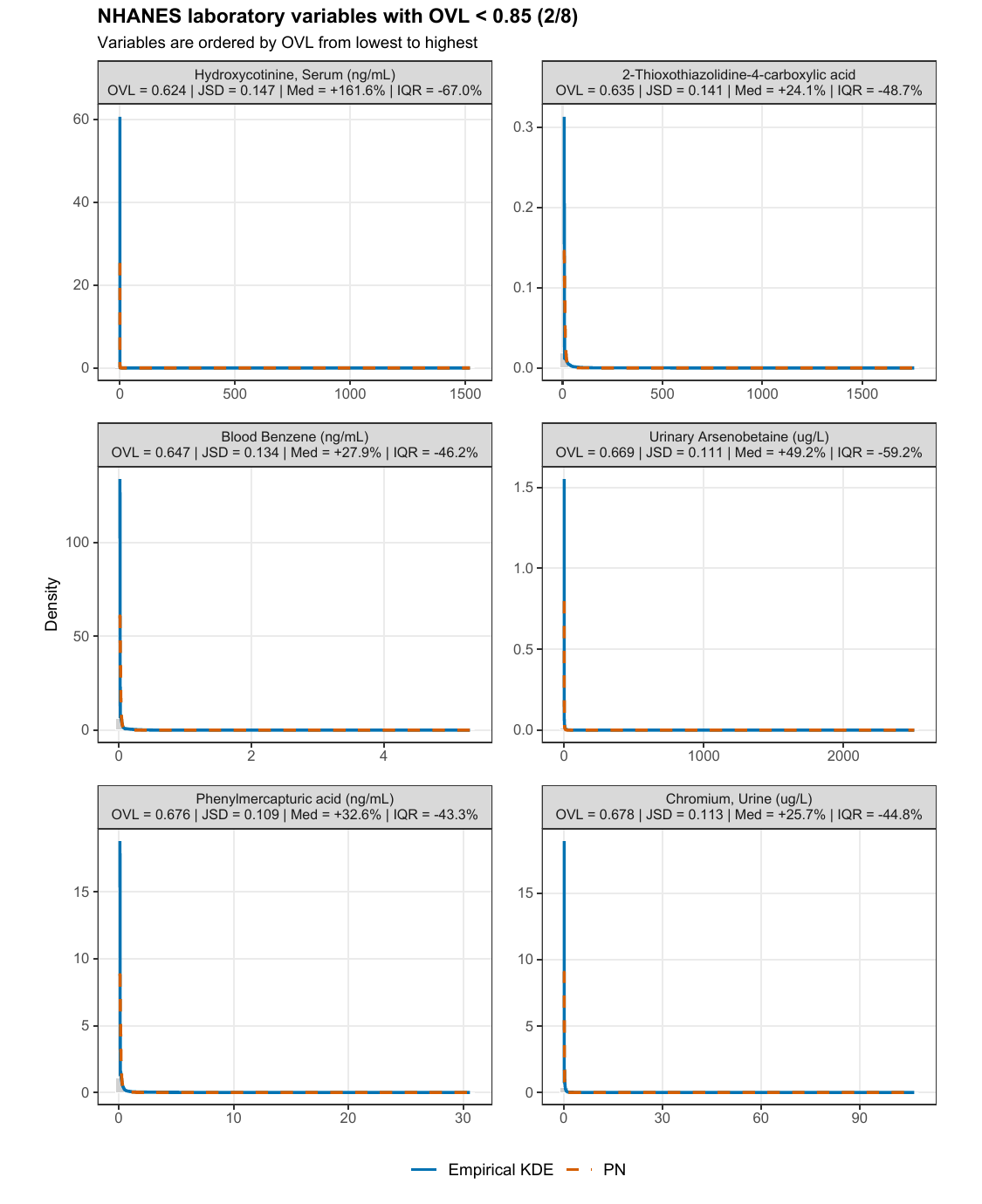}%
  {Empirical distributions and fitted PND densities on the original scale
  (continued).}
\continuedatlaspage{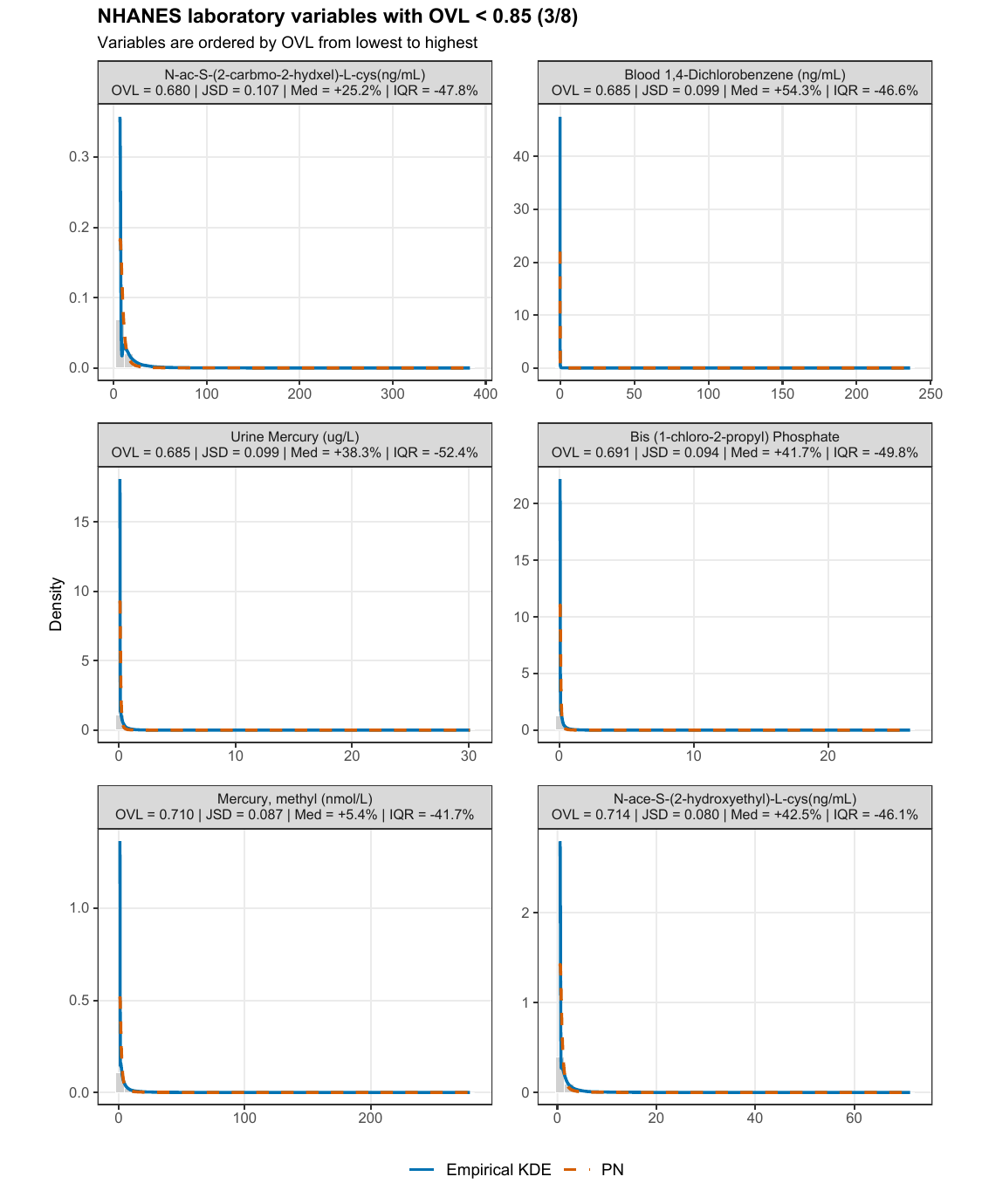}%
  {Empirical distributions and fitted PND densities on the original scale
  (continued).}
\continuedatlaspage{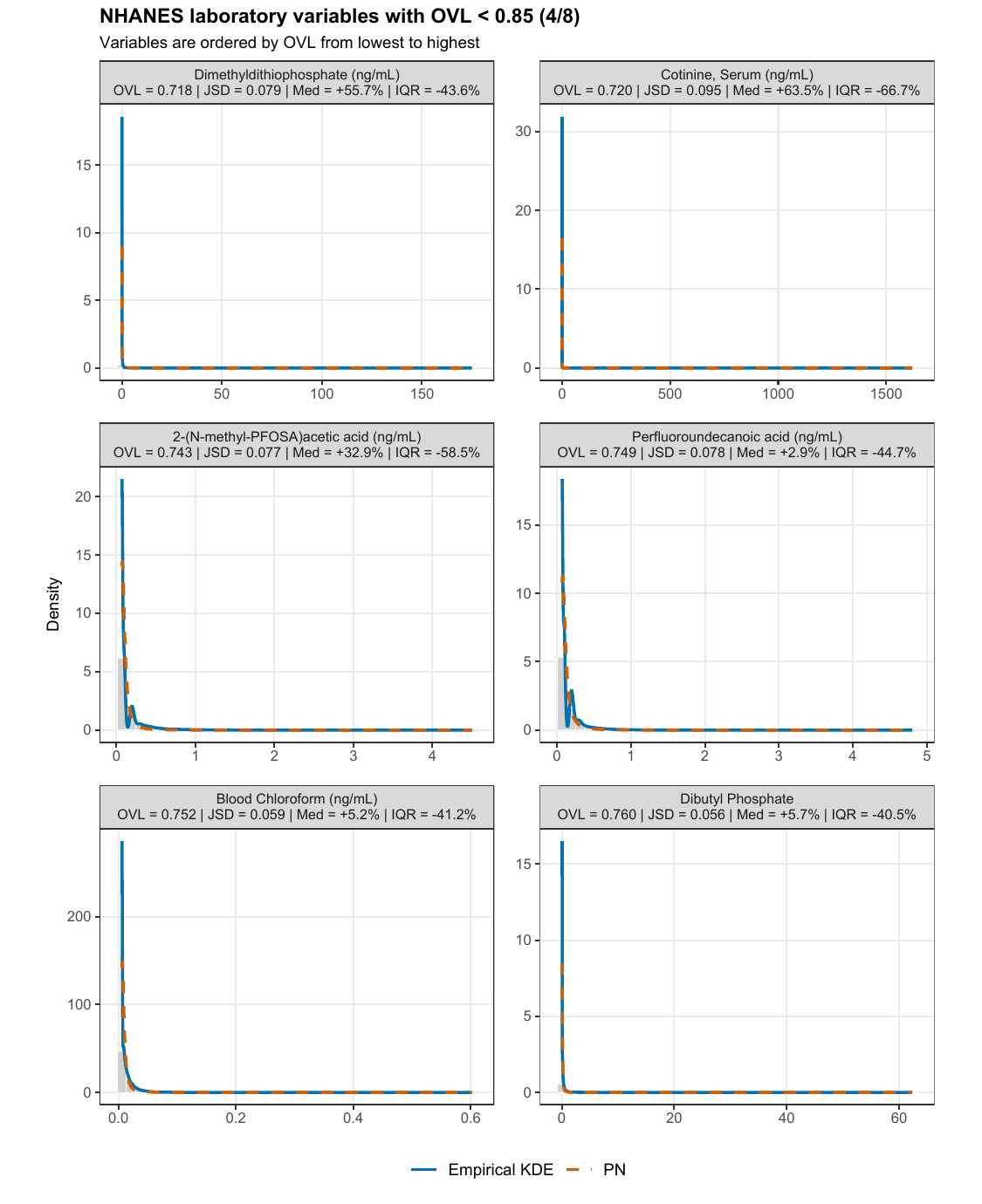}%
  {Empirical distributions and fitted PND densities on the original scale
  (continued).}
\continuedatlaspage{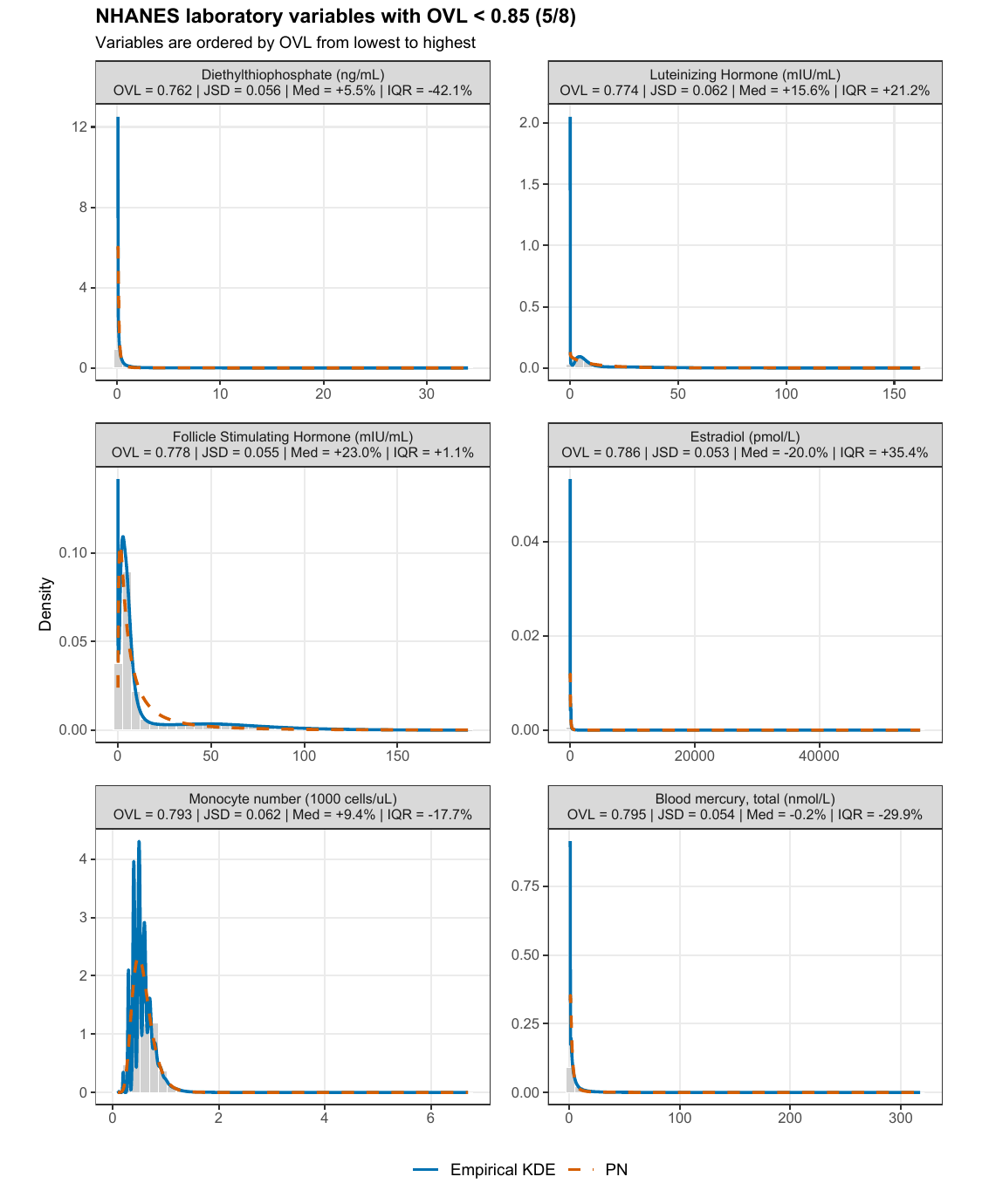}%
  {Empirical distributions and fitted PND densities on the original scale
  (continued).}
\continuedatlaspage{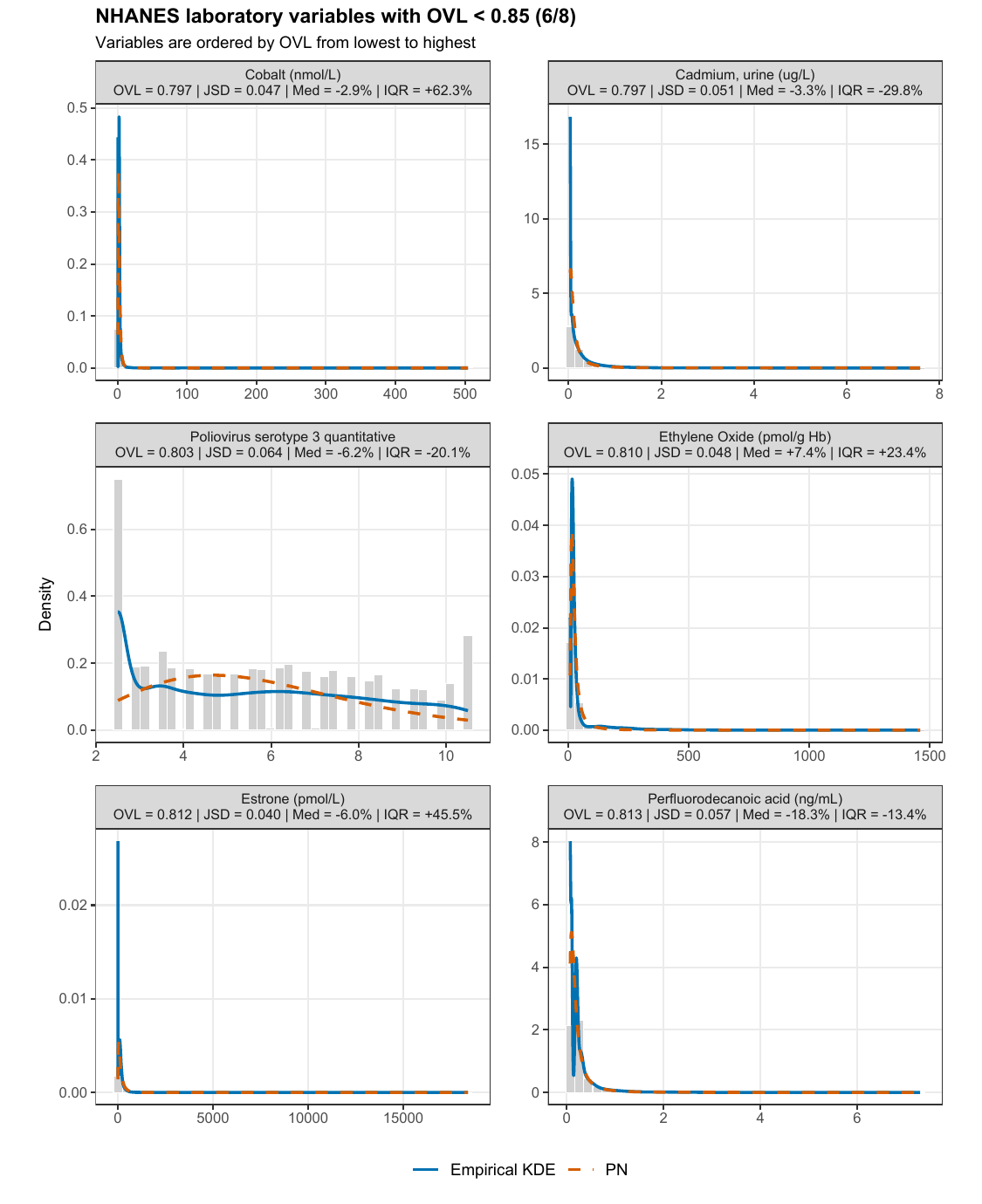}%
  {Empirical distributions and fitted PND densities on the original scale
  (continued).}
\continuedatlaspage{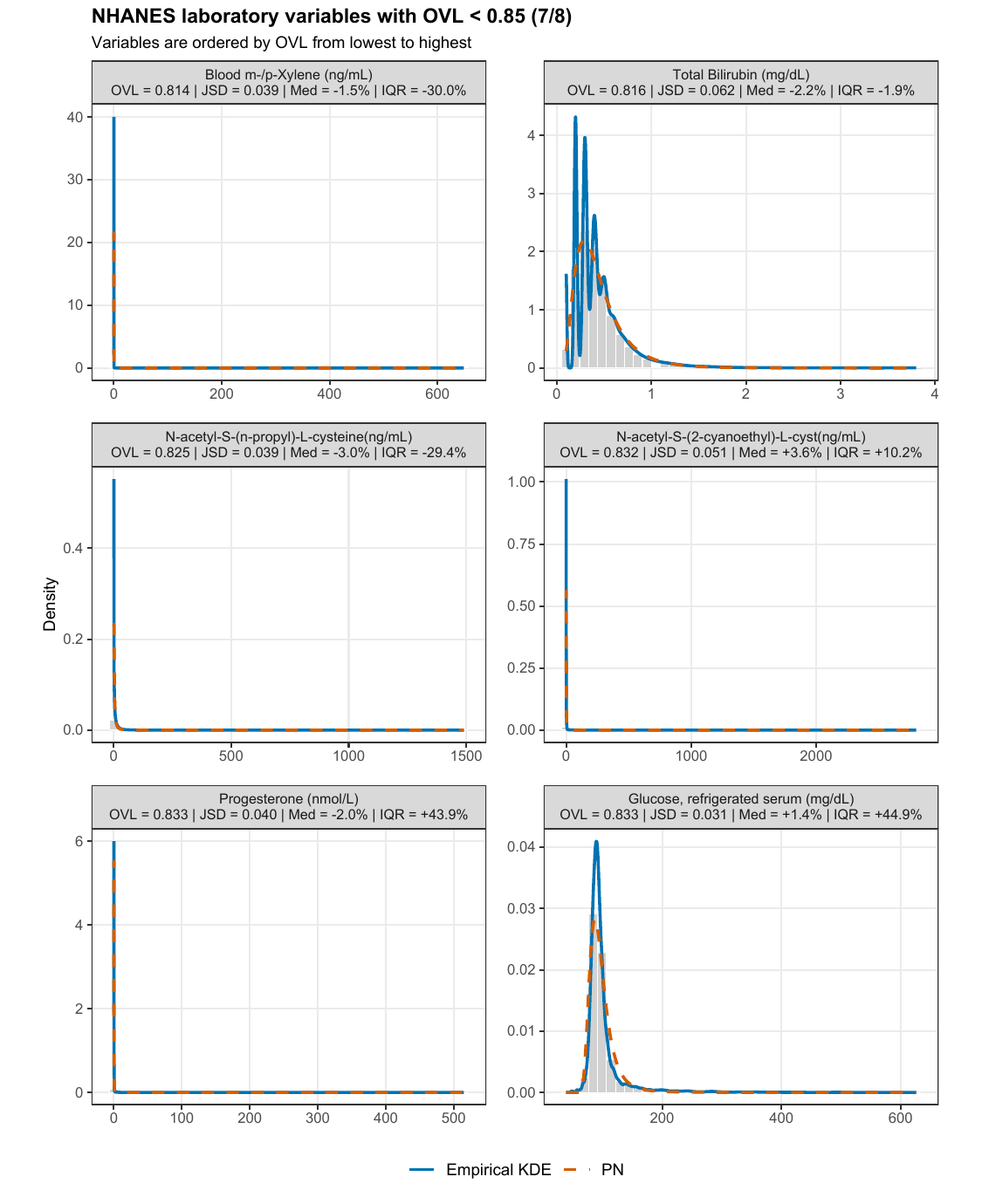}%
  {Empirical distributions and fitted PND densities on the original scale
  (continued).}
\continuedatlaspage{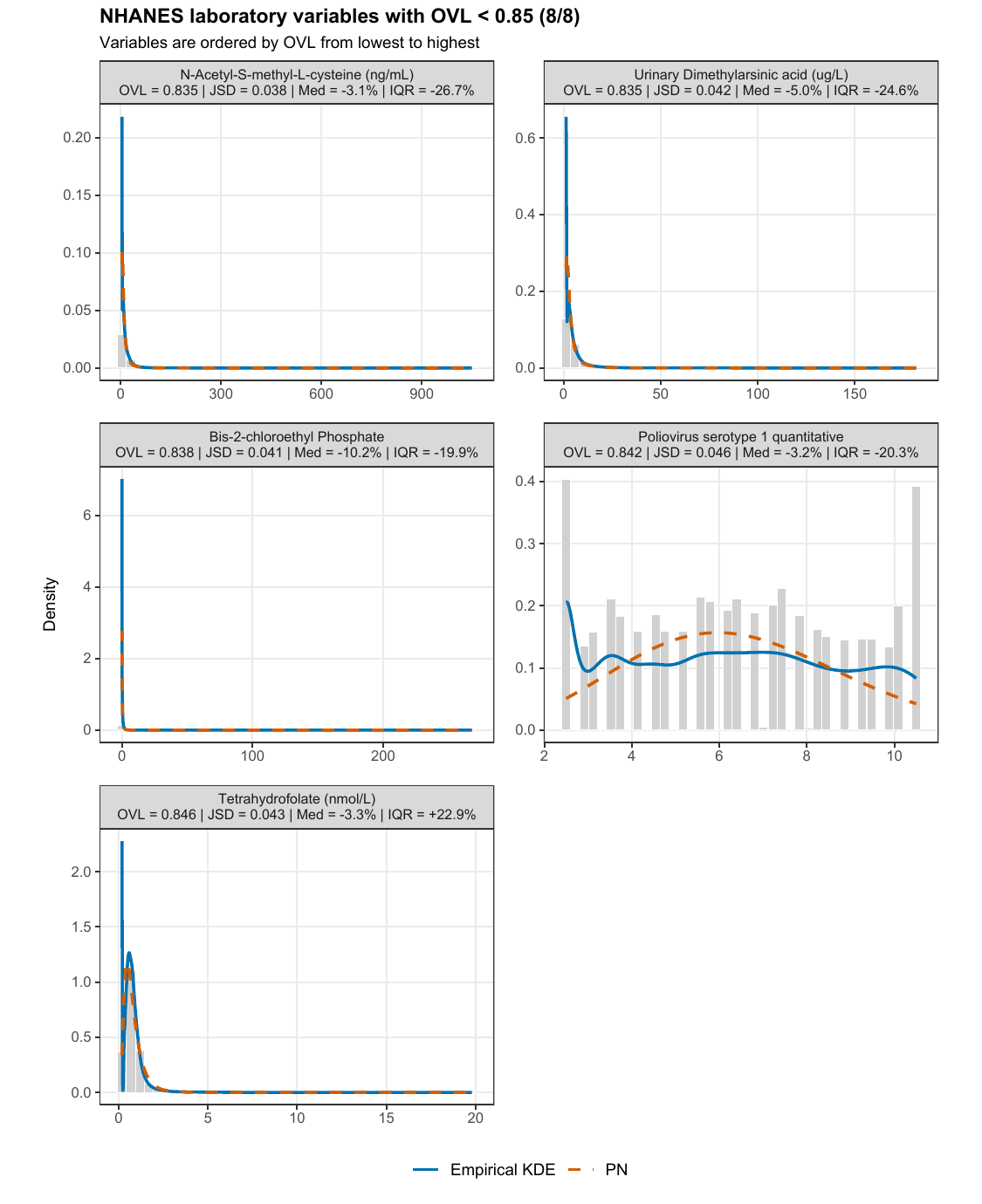}%
  {Empirical distributions and fitted PND densities on the original scale
  (continued).}


\begin{figure}[p]
  \centering
  \includegraphics[
    width=\textwidth,
    height=0.80\textheight,
    keepaspectratio
  ]{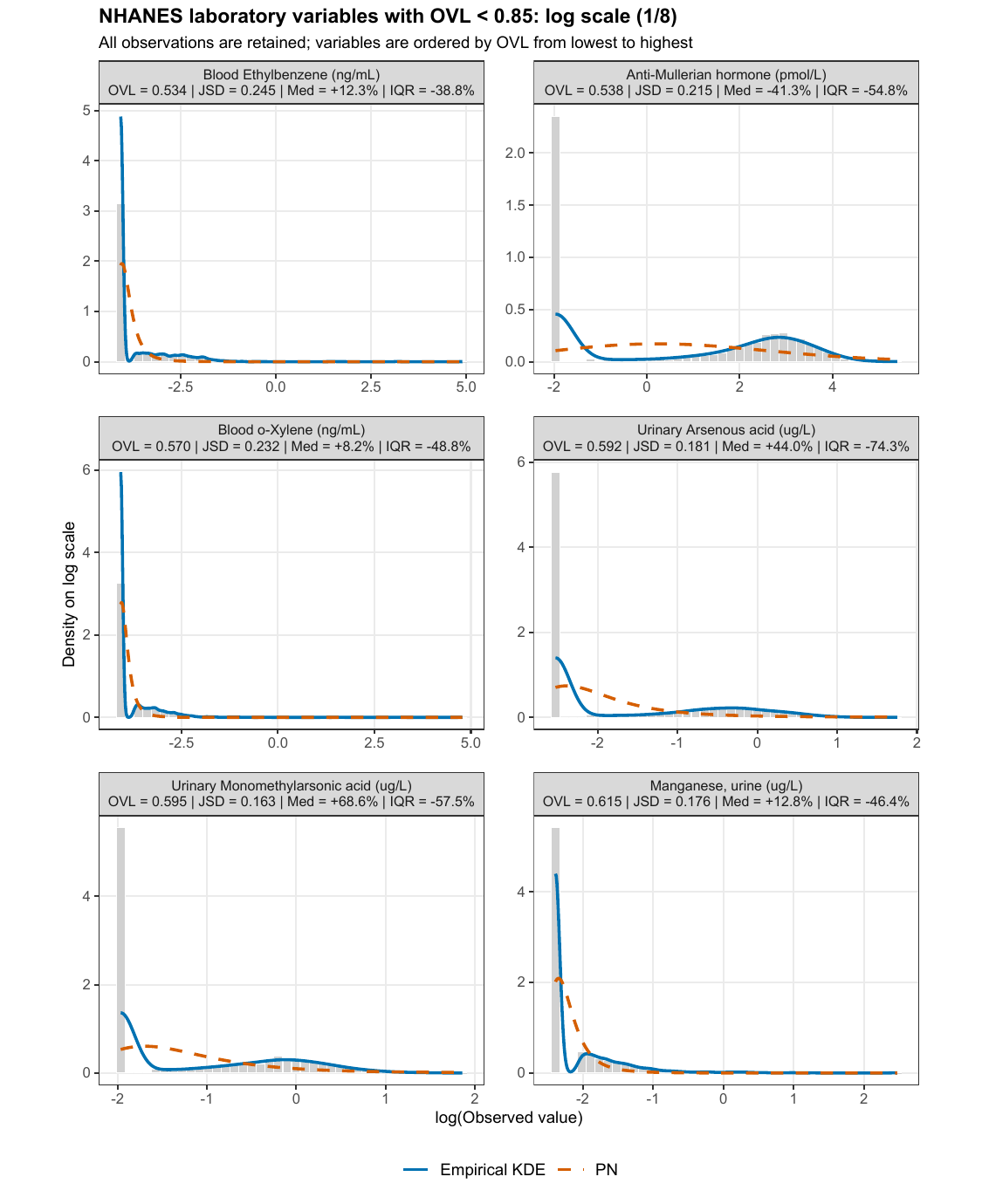}
  \caption{Empirical distributions and fitted PND densities on the
  logarithmic scale for the 47 NHANES laboratory variables with
  $\mathrm{OVL}<0.85$. Variables are shown in the same order as in
  Figure~\ref{fig:s3}. Gray bars show the empirical histogram on the
  logarithmic scale, the blue solid curve shows the Gaussian kernel density
  estimate, and the vermilion dashed curve shows the fitted PND density after
  transformation to the logarithmic scale. All observations were retained.}
  \label{fig:s4}
\end{figure}
\clearpage

\continuedatlaspage{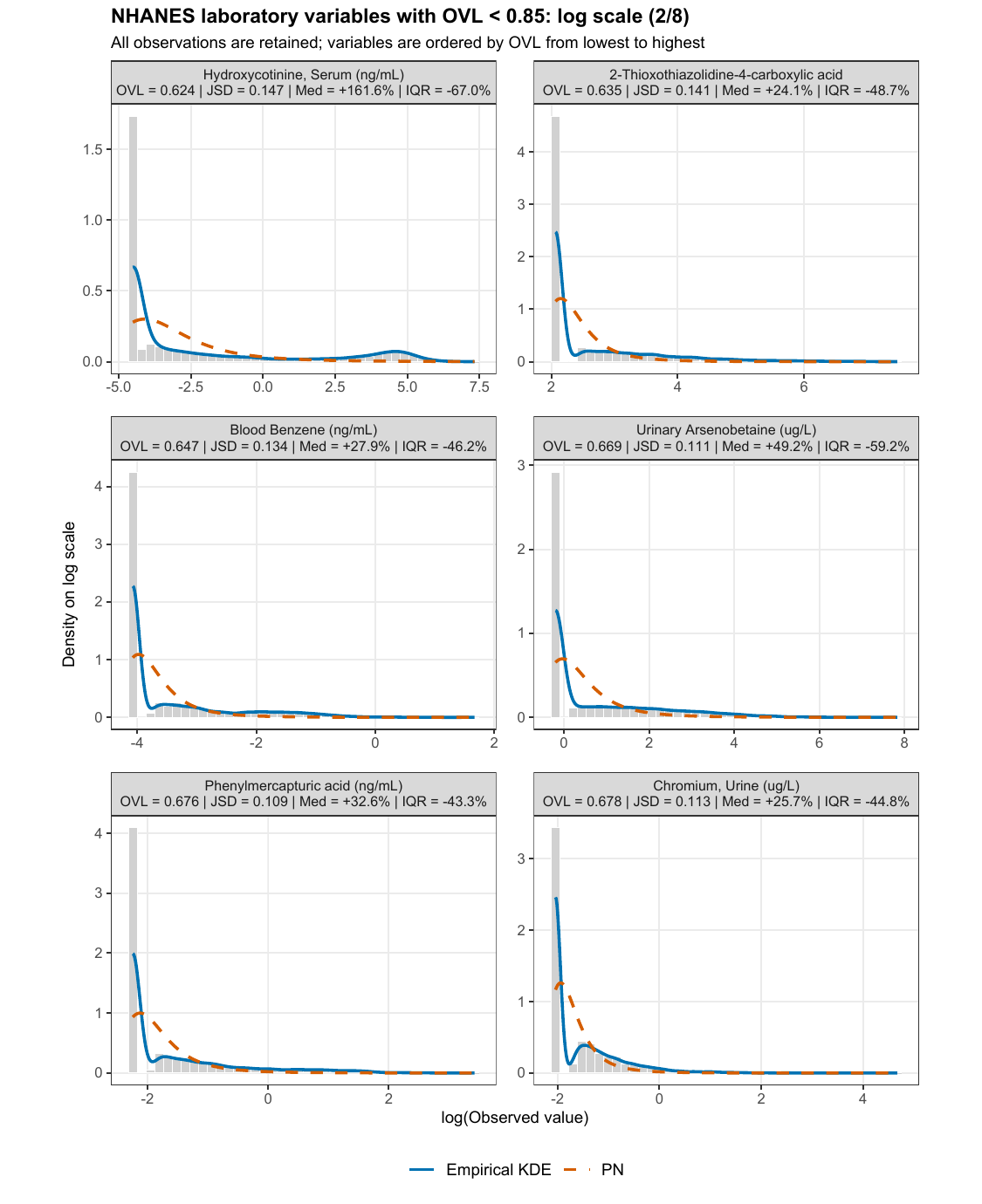}%
  {Empirical distributions and fitted PND densities on the logarithmic scale
  (continued).}
\continuedatlaspage{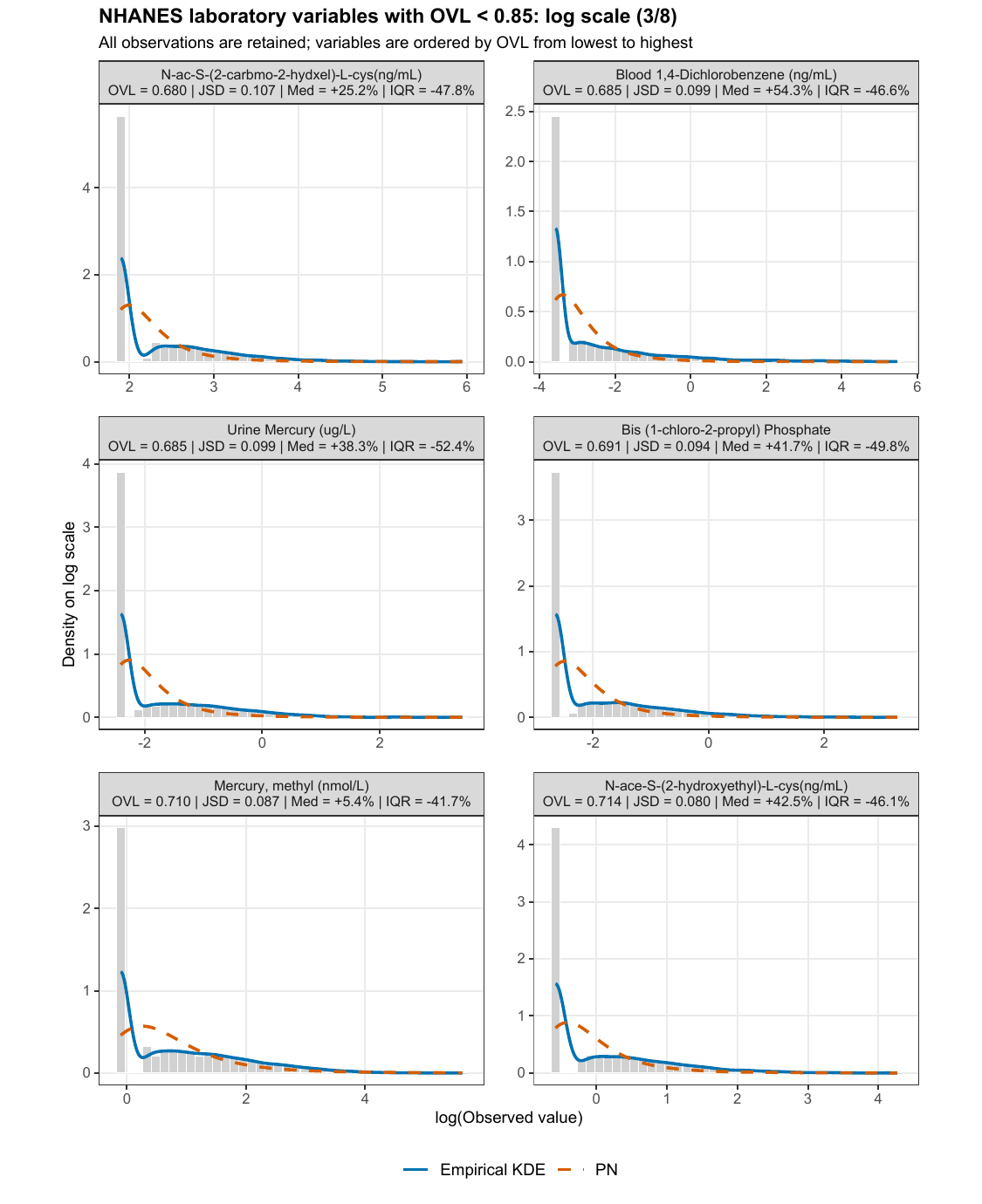}%
  {Empirical distributions and fitted PND densities on the logarithmic scale
  (continued).}
\continuedatlaspage{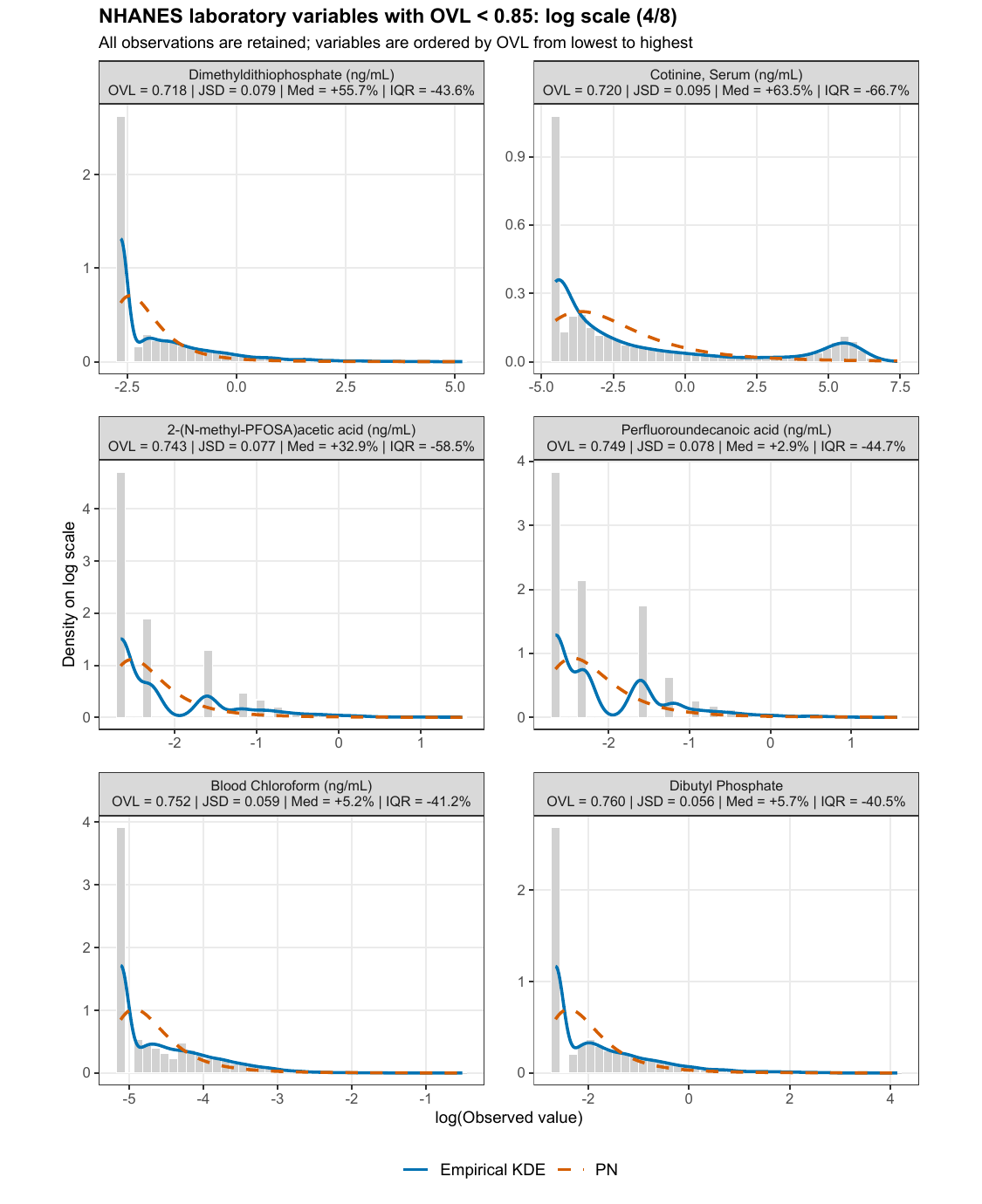}%
  {Empirical distributions and fitted PND densities on the logarithmic scale
  (continued).}
\continuedatlaspage{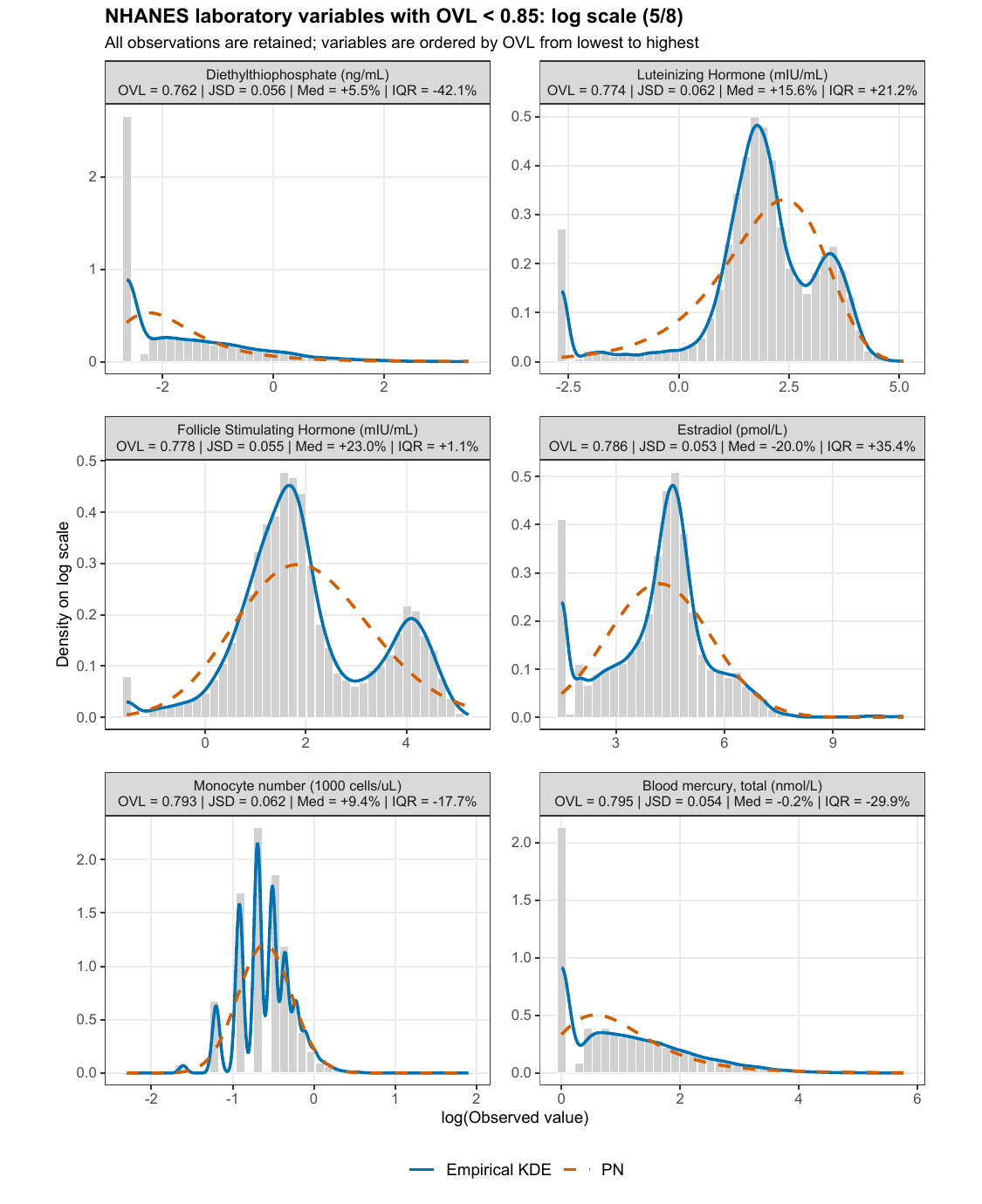}%
  {Empirical distributions and fitted PND densities on the logarithmic scale
  (continued).}
\continuedatlaspage{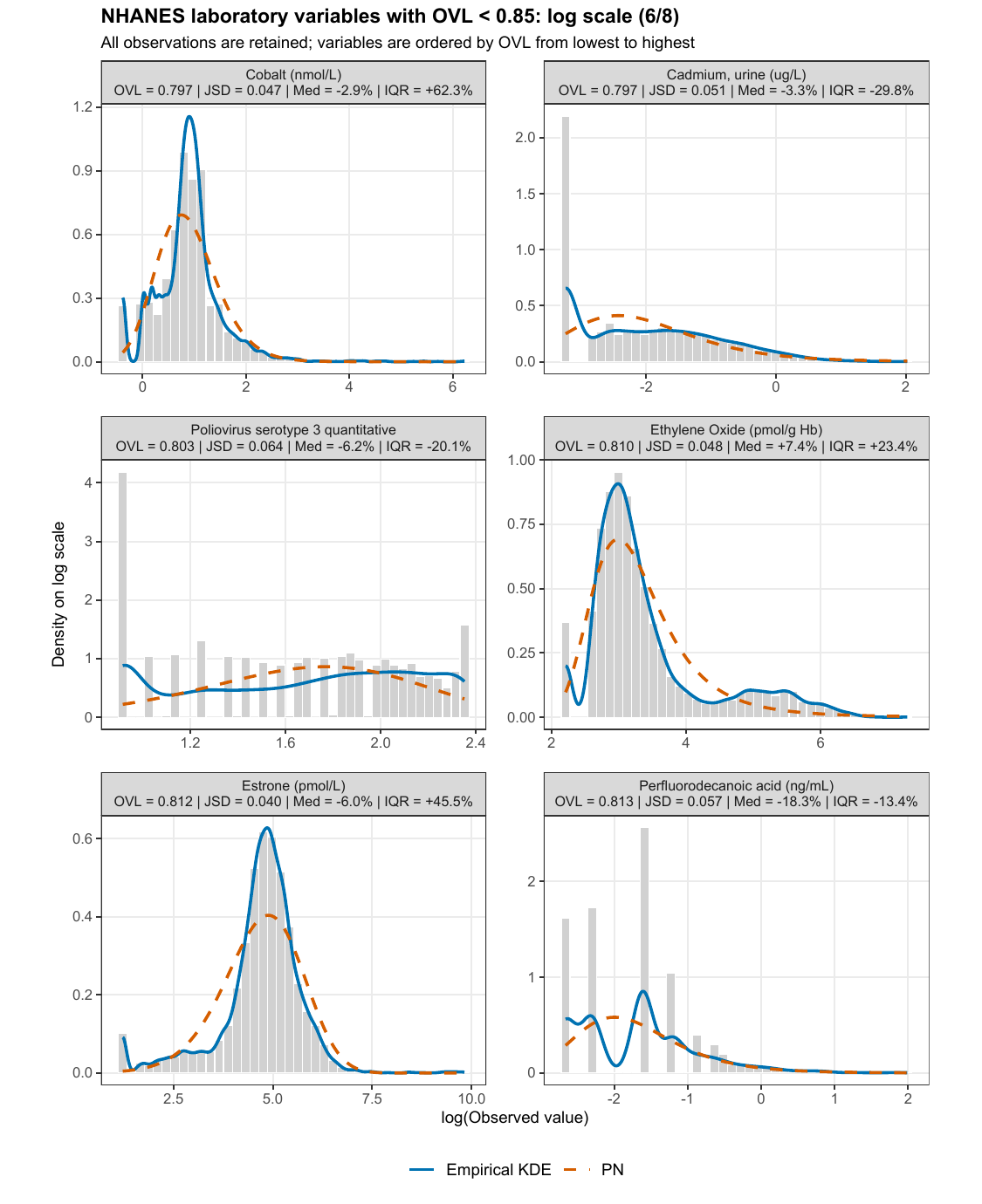}%
  {Empirical distributions and fitted PND densities on the logarithmic scale
  (continued).}
\continuedatlaspage{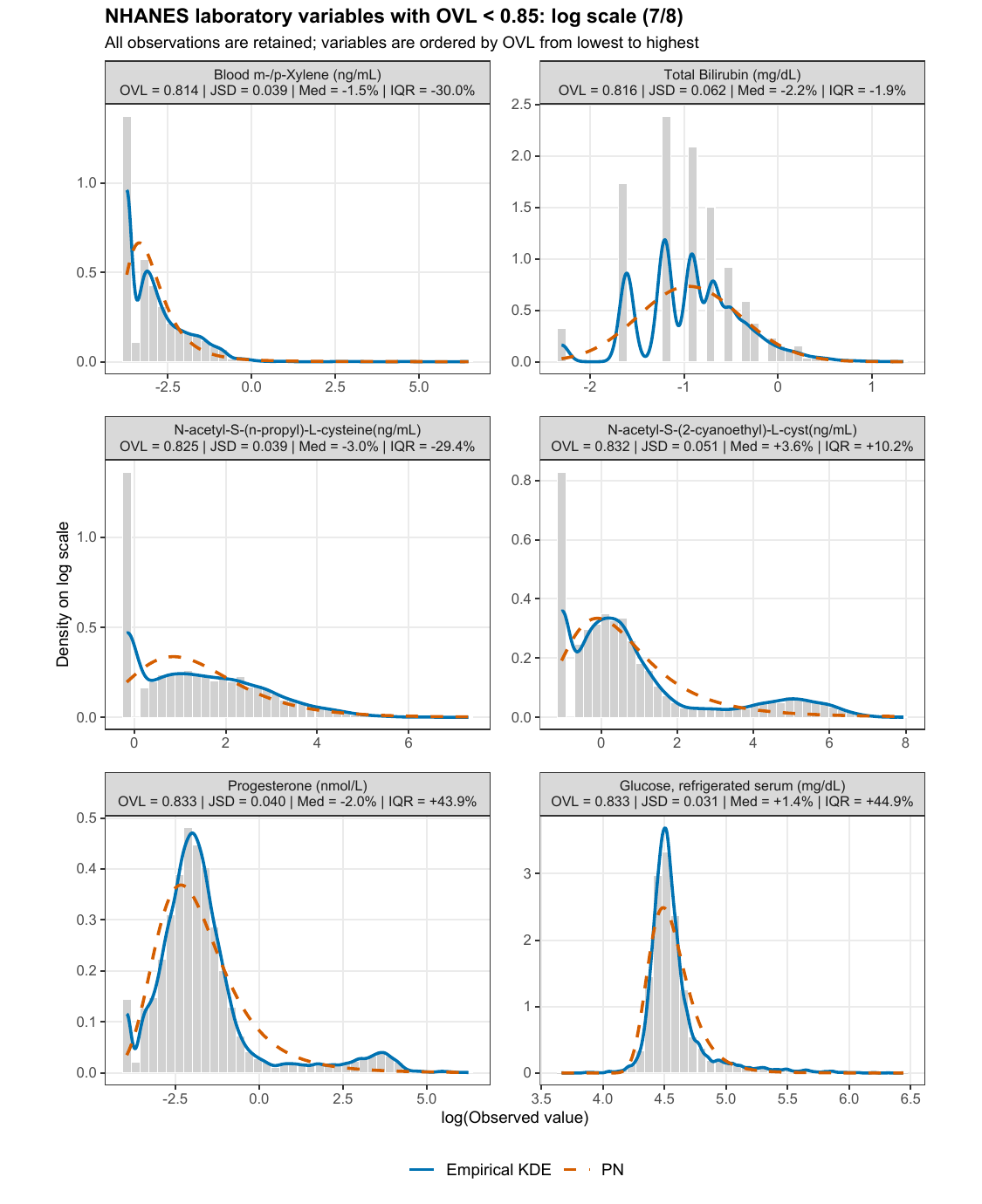}%
  {Empirical distributions and fitted PND densities on the logarithmic scale
  (continued).}
\continuedatlaspage{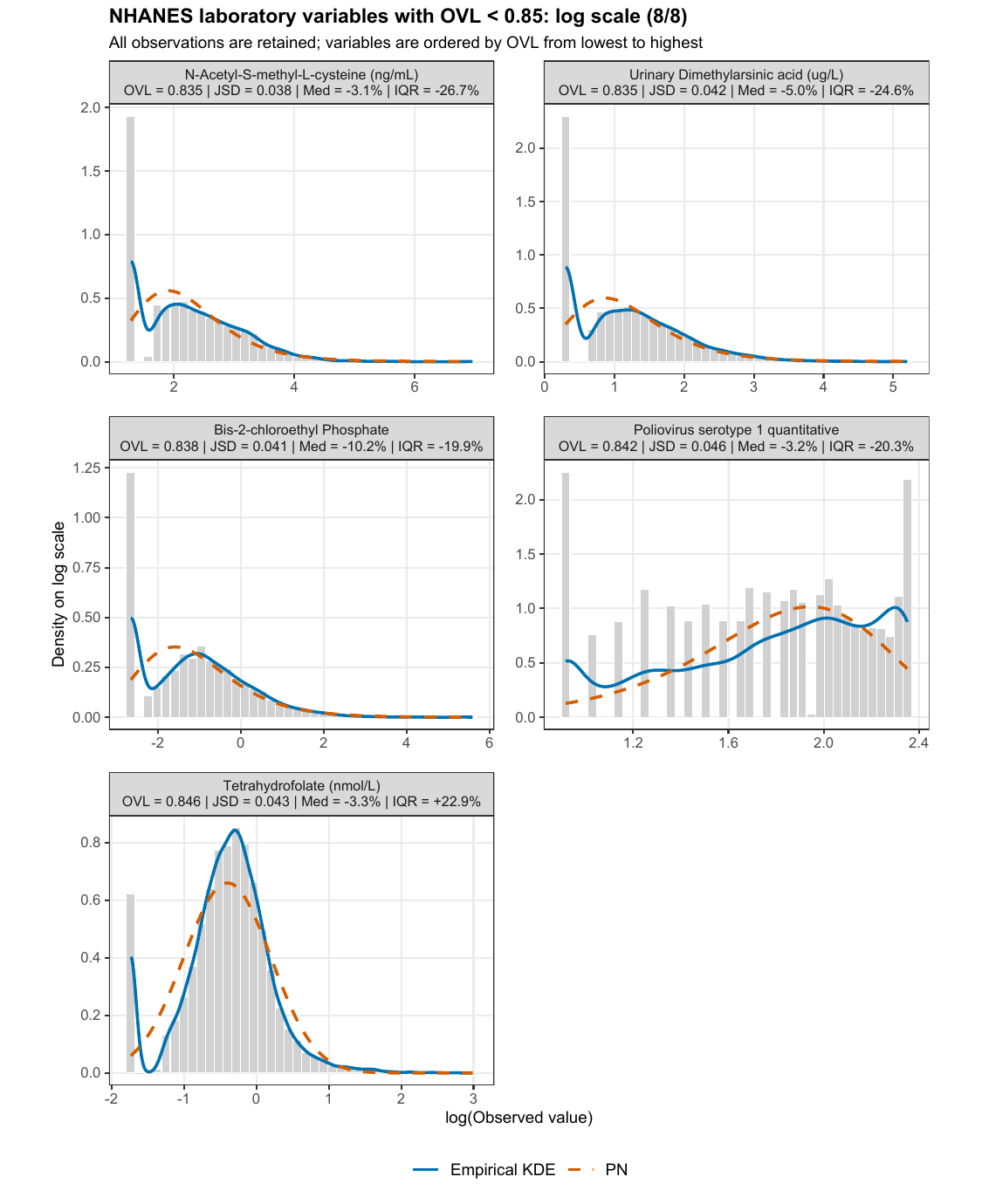}%
  {Empirical distributions and fitted PND densities on the logarithmic scale
  (continued).}

\end{document}